\documentclass[runningheads]{llncs}

\usepackage{amsmath}[2000/07/18]

\usepackage{hyperref}
\usepackage{xparse}
\usepackage{citecmds}
\usepackage{soul}
\usepackage[utf8]{inputenc}
\usepackage{float}
\usepackage{booktabs}

\usepackage{siunitx}
\usepackage[capitalise]{cleveref}

\crefname{algorithm}{Algorithm}{algorithms}
\crefname{section}{Section}{sections}
\crefname{theorem}{Theorem}{theorems}

\newenvironment{proof-sketch}[1][\unskip]{\noindent \textit{Proof sketch #1 (full proof in
appendix). }}{\hfill}

\usepackage{dsfont}
\let\mathbb\mathds

\usepackage{pseudo}

\pseudoset{
    topsep = 0pt,
    line-height = 1.2,
}

\usepackage{tcolorbox}
\tcbuselibrary{skins,theorems}
\newtcbtheorem[
        crefname = {Algorithm}{algorithms}
    ]{procedure}{Algorithm}{pseudo/ruled, top = 1.5 \belowrulesep, bottom = 1
    \aboverulesep, pseudo/init=\parskip .2\baselineskip}{alg}

\NewDocumentCommand \InputLabel  { } {\textbf{Input:}}
\NewDocumentCommand \OutputLabel { } {\textbf{Output:}}

\NewDocumentCommand \Preamble { m m } {
    \InputLabel\enskip #1
    \par
    \OutputLabel\enskip #2
}
\usepackage{pgfplots}
\pgfplotsset{compat=1.17}
\pgfqkeys{/pgf}{number format/1000 sep={\,}}
\usepgfplotslibrary{statistics}
\usetikzlibrary{patterns}
\usetikzlibrary{calc}

\RequirePackage[inline,loadonly]{enumitem}

\newlist{inl}{enumerate*}{1}
\setlist[inl]{
    label=\mbox{\rm($\roman*$)},
}

\title{Maximin Shares Under Cardinality Constraints\thanks{
        A preliminary version of this paper appeared at AAMAS 2022 as an
        extended abstract \citep{hummel_guaranteeing_2022}.
}}

\author{Halvard Hummel\inst{1}\orcidID{0000-0001-5691-8177} \and
Magnus Lie Hetland\inst{1}\orcidID{0000-0003-4204-2017}}
\authorrunning{H. Hummel and M. L. Hetland}

\institute{
    Norwegian University of Science and Technology, Trondheim, Norway,\\
    \email{halvard.hummel@ntnu.no, mlh@ntnu.no}
}

\begin{document}
\maketitle

\begin{abstract}
    We study the problem of fair allocation of a set of indivisible items among
    agents with additive valuations, under cardinality constraints. In this
    setting, the items are partitioned into categories, each with its own limit
    on the number of items it may contribute to any bundle.
    We consider the
    fairness measure known as the \textit{maximin share} (MMS)
    \textit{guarantee}, and propose a novel polynomial-time algorithm for
    finding $1/2$-approximate MMS allocations for goods---an improvement from
    the previously best available guarantee of $11/30$. For single-category
    instances, we show that a modified variant of our algorithm is guaranteed to
    produce $2/3$-approximate MMS allocations.  Among various other existence
    and non-existence results, we show that a $(\sqrt{n}/(2\sqrt{n} -
    1))$-approximate MMS allocation always exists for goods. For chores, we show
    similar results as for goods, with a $2$-approximate algorithm in the
    general case and a $3/2$-approximate algorithm for single-category
    instances. We extend the notions and algorithms related to \textit{ordered}
    and \textit{reduced instances} to work with cardinality constraints, and
    combine these with \textit{bag filling} style procedures to construct our
    algorithms.

\end{abstract}

\section{Introduction}

The problem of fair allocation is one that naturally occurs in many real-world
settings, for instance when an inheritance is to be divided or limited resources are
to be distributed. For a long time, the research in this area primarily focused
on the allocation of divisible items, but lately the interest in the more
computationally challenging case of indivisible items has seen a surge.
(\citeauthor{bouveret_fair_2016} provide a somewhat recent overview
\cite{bouveret_fair_2016}). For this variant of the problem, many of the central
fairness measures in the literature on divisible items, such as
\textit{envy-freeness} and \textit{proportionality}, are less useful. Instead,
relaxed fairness measures, such as the \textit{maximin share} (MMS)
\textit{guarantee} \cite{budish_combinatorial_2011}, have been introduced, where
all agents receive at least as much as if they partitioned the items but were
the last to select a bundle. It is not always possible to find an MMS allocation
\cite{feige_tight_2021,kurokawa_when_2016,procaccia_fair_2014}, but good
approximations exist \cite{garg_improved_2020,ghodsi_fair_2018}.

Fairly allocating items in the real world often involves
placing constraints on the bundles allowed in an allocation. For example,
consider the problem where a popular physical conference or convention offers
a variety of talks and panels organized across several synchronized parallel
tracks. Due to space constraints, each talk is limited to some maximum number of
participants, fewer than the total number of participants at the conference.
Consequently, there may be more people interested in attending some talks than
there are available seats. To mitigate this, the conference wants to fairly
allocate the available seats, based on participants' preferences, so that no
participant receives seats they cannot use, i.e., multiple seats at the same
talk or seats at multiple talks in the same time slot. In order to solve this
problem, we need to be able to express that some items belong to the same
category (seats at talks in the same time slot) and that there is a limit on the
number of items each category can contribute to any bundle (in this case~1).
This kind of constraints is called \textit{cardinality constraints} and was
introduced by \citet{biswas_fair_2018}.

The conference example highlights a general type of problems for which
cardinality constraints are useful, where each agent should not receive more
items of a certain type than she could possibly have use for. Another such
problem is the motivating example of \citet{biswas_fair_2018}: A museum is to
fairly allocate exhibits of different types to newly opened branches. To make
sure that each branch can handle its allocated exhibits, so that no exhibits go
to waste, an upper limit is placed on the number of exhibits each branch can be
allocated of each exhibit type. The constraints may also provide each agent with
some diversity in the type of items she receives. For example, with sufficiently
small limits in the museum example, each branch must receive a somewhat diverse
collection of exhibits.

Another application is making sure that
items of certain types are guaranteed to be roughly evenly distributed among the
agents. This can be achieved by setting the number of items each agent can
receive from a given category close to the number of items in this category
divided by the number of agents. For example, consider a situation where a set
of donated items, including a limited number of internet-capable devices, are to
be fairly allocated to low-income families. A single family can make use of many
internet-capable devices. However, the organization behind the allocation
process may want to make sure that as many families as possible have access to
the internet. By placing all the internet-capable devices in the same category
and giving each family at most one item from this category, the internet-capable
devices will
be distributed to as many families as possible.

\Citet{biswas_fair_2018} showed that under cardinality constraints,
with additive valuations, it is always possible to find an allocation of goods
where each agent gets at least $1/3$ of her MMS. This is achieved by a reduction
to an unconstrained setting with submodular valuations, where the approximate
allocation is found using an algorithm described by \citet{ghodsi_fair_2018}.
More recently, \citeauthor{Li:2021} showed that $11/30$-approximate MMS
allocations are guaranteed to exist under hereditary set system constraints
\cite{Li:2021}. This approximation guarantee is achievable in polynomial time
for certain classes of set systems, including set systems representing
cardinality constraints.

\subsection{Contributions}

We develop a polynomial-time algorithm for finding $1/2$-approximate MMS
allocations for goods under cardinality constraints, improving on the $1/3$ and
$11/30$ guarantees of \citeauthor{biswas_fair_2018} \cite{biswas_fair_2018} and
\citeauthor{Li:2021} \cite{Li:2021}, which are, to our knowledge, the best
guarantees previously available. To construct the algorithm, we extend the
notions and algorithms related to \emph{ordered} and \emph{reduced instances} to
work with cardinality constraints, and combine these with a \emph{bag-filling}
style algorithm. Combining
this
algorithm with a lone-divider
style \cite{Aigner-Horev:2022} preprocessing step, we show that
$(\sqrt{n}/(2\sqrt{n} - 1))$-approximate MMS allocations always exist for
goods---a large improvement for few agents. The preprocessing step unfortunately
relies on finding MMS-partitions, an NP-hard problem
\cite{woeginger_polynomial-time_1997}. However, the $1/2$-approximate MMS
algorithm
is
able to find both $(n/(2n - 1))$-approximate MMS
allocations and $1$-out-of-$(2n - 1)$ MMS allocations by changing a constant.

For chores, we show that a similar approach finds $2$-approximate
(or, more precisely, $((2n - 1)/n)$-approximate)
MMS
allocations in polynomial time.
This is, to our knowledge, the first MMS result for chore
allocation under cardinality constraints.

We also examine a special case of cardinality constraints, in which all the
items belong to the same category. This case is equivalent to placing a
restriction on the number of items in each bundle, or equivalently restricting
bundles to independent sets of a \emph{uniform matroid}. This is a setting of
interest in itself, especially for chores, where it can be useful to make sure
that no agent is stuck with a much larger number of chores than anyone else. By
modifying our general algorithms, we show that in this special case,
$(2/3)$-approximate MMS allocations for goods and $(3/2)$-approximate MMS
allocations for chores can be found in polynomial time.

\subsection{Related Work}

Several other constraint types have been examined in the recent literature.
(See \citeauthor{Suksompong:2021}'s recent survey for a detailed overview
\cite{Suksompong:2021}.) One such constraint is that all agents must receive
exactly the same number of items \cite{ferraioli_regular_2014}, a more
restrictive version of our single-category instances. Another, studied by
\citeauthor{bouveret_fair_2017}, uses an underlying graph to represent
connectivity between the items and requires each bundle to form a connected
component \cite{bouveret_fair_2017}. Such connectivity constraints have since
been explored in many papers
\citep[e.g.,][]{bilo_almost_2018,greco_complexity_2020,lonc_maximin_2018}.  A
variation is the allocation of \emph{conflicting items}, where each bundle must be
an independent set in the graph~\citep{chiarelli_packing_2020,hummel_fair_2021}.
There is some overlap between this scenario and cardinality constraints with
threshold~1~\citep[cf.][]{hummel_fair_2021}, but neither is a generalization of
the other. Cardinality constraints have recently been studied by
\citeauthor{Shoshan:2022}, who considered the problem of finding allocations
that are both Pareto optimal and EF1 for instances with two agents
\cite{Shoshan:2022}.

Matroids have been used to constrain allocations in several different ways
\cite{gourves_near_2014}. The cardinality constraints placed on a single bundle
may in fact be represented by a \emph{partition matroid} or for single-category
instances a \emph{uniform matroid}. The $1/2$-approximate MMS algorithm of
\citet{gourves_maximin_2019} applies to the superficially similar problem where
a single matroid constraint is placed on the union of \emph{all} bundles. As
pointed out by \citet{biswas_fair_2018}, this algorithm cannot be applied to the
cardinality constraint scenario.

\section{Preliminaries}\label{sec:preliminaries}

For a given instance, $I = \langle N, M, V \rangle$, of the fair allocation
problem, let $N = \{1, 2, \dots, n\}$ denote a set of \textit{agents}, $M =
\{1, 2, \dots, m\}$, a set of \textit{items}, and
$V = \langle v_1, v_2, \dots, v_n \rangle$, the \emph{valuation profile},
i.e., the collection of the agents' \emph{valuation functions}
$v_i : 2^m \to \mathbb{R}$ over the
subsets $S \subseteq M$. For simplicity, the valuation of a single item
$v_i(\{j\})$ will be denoted by both $v_i(j)$ and $v_{ij}$. We assume that the
valuations are additive, i.e., $v_i(S) = \sum_{j \in S} v_{ij}$.
We
wish to find an allocation $A =
\langle A_1, A_2, \dots, A_n \rangle$ that forms a partition of $M$ into $n$
possibly empty subsets, or \emph{bundles}, one for each agent. We say that an
instance $I$ consists of \emph{goods} if $v_{ij} \ge 0$ for all $i \in N, j \in
M$, and \emph{chores} if $v_{ij} \le 0$ for all $i \in N, j \in M$.
We consider both instances consisting of goods and ones consisting of
chores. However, we do not consider instances consisting of a mix of goods and
chores. For simplicity, we will throughout the paper assume that all instances
consist of goods, except for in \cref{sec:chores}, which covers our results on
chores.

For the fair allocation problem \emph{under cardinality constraints}, an
instance is given by $I = \langle N, M, V, C \rangle$, where $C$ is a set of
$\ell$ pairs $\langle C_h, k_h \rangle$ of \emph{categories} $C_h$ and
corresponding \emph{thresholds} $k_h$. The categories constitute a partition of
the items, $M$. An allocation $A$ is \textit{feasible} for the instance if no
agent receives more than $k_h$ items from any category $C_h$, i.e., if $|A_i
\cap C_h| \leq k_h$ for all $i \in N, h \in \{1, \dots, \ell\}$. We let
$\mathcal{F}_I$ denote the set of all feasible allocations for $I$, with the
subscript omitted if it is clear from context. To guarantee that there is at
least one feasible allocation, i.e., $\mathcal{F} \neq \emptyset$, no category
may contain more items than we can possibly distribute, i.e., we require that
$|C_h| \le nk_h$ for all $h \in \{1, \dots, \ell\}$\footnote{Instances with more
than $nk_h$ items in category $C_h$ can be handled by ordering the instance
(see \cref{sec:ordered}) and ignoring the worst items in the category.}.

We are concerned with the fairness criterion known as the \textit{maximin share
guarantee} \cite{budish_combinatorial_2011}. The \emph{maximin share} (MMS) of
an agent is the value of the most preferred bundle the agent can guarantee
herself if she were to divide the items into feasible bundles and then choose
her own bundle last. More formally:\footnote{The definition is equivalent for
chores.}

\begin{definition}
    Let $I = \langle N, M, V, C \rangle$ be an instance of the fair allocation
    problem under cardinality constraint. The \emph{maximin share} of an agent
    $i$ for the instance $I$ is given by
    \begin{equation*}
        \mu^I_i = \max_{A \in \mathcal{F}_I} \min_{A_j \in A} v_i(A_j)\,,
    \end{equation*}
    where $\mathcal{F}_I$ is the set of feasible allocations for $I$.
    If $I$ is obvious from context, we write simply $\mu_i$.
\end{definition}

\noindent An allocation is said to \emph{satisfy the MMS guarantee}, or \emph{to
be an MMS allocation}, if each agent gets a bundle valued at least as much as
the agent's MMS, i.e., $v_i(A_i) \ge \mu_i$ for all agents $i$. We concern
ourselves with allocations that satisfy this guarantee \emph{approximately},
where an allocation is said to be an \emph{$\alpha$-approximate MMS allocation}
for some $\alpha>0$ if $v_i(A_i) \ge \alpha\mu_i$ for all agents $i$. An
allocation $A$ is said to be an \emph{MMS partition} of an agent $i$, if $v_i(A_j)
\ge \mu_i$ for all $A_j \in A$. By definition, at least one MMS partition exists
for any agent in any instance. As MMS allocations are not guaranteed to exist
\cite{feige_tight_2021,kurokawa_when_2016,procaccia_fair_2014}, there exists a
generalized and relaxed version of MMS, called the \emph{$l$-out-of-$d$ MMS}
\cite{Babaioff:2021}.%
\footnote{We use $l$ instead of the usual $\ell$ to avoid conflicting use of
symbols.}
This fairness criterion works like MMS, except that the agent is to partition
the goods into $d$ feasible bundles maximizing the combined value of the $l$
least valuable bundles in the partition.
Our algorithms require some knowledge about the value of $\mu_i$ in order to
determine
when a bundle is worth at least $\alpha\mu_i$ to an agent $i$. Finding the MMS
of an agent is known to be NP-hard for the unconstrained fair allocation problem
\cite{woeginger_polynomial-time_1997}. Since unconstrained fair allocation is
simply the special case of $\ell = 1$ and $k_1 = m$, finding an agent's MMS is
at least as hard under cardinality constraints.\footnote{In the unconstrained
setting, a PTAS exists for finding the MMS of each individual agent
\cite{woeginger_polynomial-time_1997}, but this PTAS does not extend to fair
allocation under cardinality constraints and there does not, to our knowledge,
exist a PTAS for this problem.} In order to provide polynomial-time algorithms,
we exploit the fact that $\mu_i$ cannot be larger than the average bundle value,
i.e., $\mu_i \le v_i(M)/n$, and we can scale all values so that $v_i(M)=n$,
so that $\mu_i\leq 1$, as shown in the following theorems.
Due to space constraints,
their proofs have been omitted, but can be found in the appendix along with all
other omitted proofs. The proofs from ordinary fair allocation for the two
succeeding theorems do in fact extend to cardinality constraints without any
modification \citep[see,
e.g.,][]{amanatidis_approximation_2017,garg_approximating_2019}.
We assume, without loss of generality, that $v_i(M)>0$ for each agent $i$.%
\footnote{If $v_i(M) = 0$,
normalization does not work. However, since this implies $\mu_i = 0$,
\cref{cor:single-item-reduction} can be used to eliminate agent $i$ from the
instance.}

\begin{theorem}[Scale invariance]
    \label{thm:scale-inv}
    If $A$ is an MMS allocation for the instance $I = \langle N, M, V, C
    \rangle$, then $A$ is also an MMS allocation for $I' = \langle N, M, V',
    C\rangle$, where $v_i'(S) = a_i v_i(S), a_i > 0$, for some agent $i$.
\end{theorem}

\begin{theorem}[Normalization]\label{thr:normalized}
    Let $I = \langle N, M, V, C\rangle$ be an instance of the fair allocation
    problem of under cardinality constraints and $v_i(M) = |N|$ for some agent
    $i$. Then $\mu_i \leq 1$.
\end{theorem}

\noindent Once valuations have been normalized, constructing an
$\alpha$-approxi\-mate MMS allocation reduces to providing each agent with a
bundle worth at least $\alpha$.

\section{Ordered Instances}\label{sec:ordered}

In the unconstrained setting, \citeauthor{bouveret_characterizing_2016} showed
that each instance can be reduced to an instance where all agents have the same
preference order over all goods \cite{bouveret_characterizing_2016}. That is, in
such an instance there exists an ordering of the goods such that when $j < k$, we
have $v_{ij} \ge v_{ik}$ for all agents $i$. While
\citeauthor{bouveret_characterizing_2016} introduced these as instances that
satisfy \textit{same-order preferences}, we will refer to them as
\textit{ordered instances}, as is the norm for MMS-approximation algorithms
\cite{barman_approximation_2017,garg_approximating_2019}.

The reduction works as follows. For each agent, sort the good values
and reassign these to the goods, which are listed in some predetermined order,
common to all agents. Allocations for the reduced instance are converted into
allocations for the original instance, without diminishing their value,
by going through the goods in the predetermined order; the agent who originally
received a given good instead chooses her highest-valued remaining good.

Since only the permutation of value assignments to goods changes,
the reduction does not change
the MMS of each agent. Thus, any $\alpha$-approximate MMS allocation in
the ordered instance will also be $\alpha$-approximate in the original
instance. Ordered instances are therefore at least as hard as any other
instances, and it suffices to show that an algorithm produces an
$\alpha$-approximate MMS allocation for ordered instances.

The standard definition of an ordered instance does not work under cardinality
constraints, due to an inherent loss of information about which goods belong to
which category. Without this information, one is not guaranteed to be able to
produce a feasible $\alpha$-approximate MMS allocation when converting back to
the original instance. We generalize the definition to fair
allocation under cardinality constraints. In the special case where $\ell = 1$,
this definition and the later conversion algorithms are equivalent to those of
\citeauthor{bouveret_characterizing_2016}.

\begin{definition}
    An instance $I = \langle N, M, V, C \rangle$ of the fair allocation problem
    under cardinality constraints is called an \emph{ordered instance} if each
    category $C_h=\{c_1,c_2,\cdots,c_{|C_h|}\}$ is ordered such that for all
    agents $i$, $v_i(c_1) \ge v_i(c_2) \ge \dots \ge v_i(c_{|C_h|})$.
\end{definition}

\noindent With the generalized definition, the reduction of MMS-approximation to
ordered instances can be extended to cardinality constraints by applying the
algorithms of \citeauthor{bouveret_characterizing_2016} to each category $C_h$
individually, as shown in \cref{alg:convert,alg:recover}.

\begin{figure}[t]
\begin{procedure}[after={}, width=.48\textwidth, equal height group=A]
    {Order instance}{convert}
\Preamble{Instance $I = \langle N, M, V, C \rangle$}
         {Ordered $I' = \langle N, M, V', C\rangle$}
\begin{pseudo}
    \kw{for} $(C_h, k_h) \in C$ \\+
        \kw{for} $j = 1$ \kw{to} $|C_h|$ \\+
            \kw{for} $i \in N$ \\+
                $v_i'(c_j) = $ $i$'s $j$th highest \\++
                                value in $C_h$ \\-----
    \kw{return} $\langle N, M, V', C\rangle$
\end{pseudo}
\end{procedure}
\hfill
\begin{procedure}[before={}, width=.48\textwidth, equal height group=A]{Recover
    solution}{recover}
\Preamble{Instance $I = \langle N, M, V, C \rangle$ and allocation $A'$ for
    corresponding $I'$}{Allocation $A$ for $I$}
\begin{pseudo}
    $A = \langle \emptyset, \dots, \emptyset \rangle$ \\
    \kw{for} $(C_h, k_h) \in C$ \\+
        \kw{for} $j = 1$ \kw{to} $|C_h|$ \\+
            $i = $ agent for which $c_j \in A_i'$ \\
            $j^* = $ $i$'s preferred item in $C_h$ \\
            $A_i = A_i \cup \{j^*\}$ \\
            $C_h = C_h \setminus \{j^*\}$ \\--
    \kw{return} $A$
\end{pseudo}
\end{procedure}
\end{figure}

\begin{lemma}\label{lem:ordered}
    Let $I = \langle N, M, V, C \rangle$ be an instance of the fair allocation
    problem under cardinality constraints, and $A'$ a feasible
    $\alpha$-approximate MMS allocation for the ordered instance $I'$ produced
    by \cref{alg:convert}. Then the allocation produced by conversion of $A'$
    with \cref{alg:recover} is a feasible $\alpha$-approximate MMS allocation
    for $I$.
\end{lemma}

\noindent Repeating the ordering and deordering procedure for each category does
not affect the polynomial nature of the procedures. As a result, the reduction
to ordered instances holds.

\begin{theorem}\label{thr:ordered}
    For fair allocation under cardinality constraints, MMS-approx\-imation
    reduces to MMS-approximation of ordered instances in polynomial time.
\end{theorem}

\begin{proof}
    By \cref{lem:ordered} it is sufficient to find an $\alpha$-approximate MMS
    allocation for the reduced instance produced by \cref{alg:convert}. Since
    both \cref{alg:convert,alg:recover} are polynomial in the number of agents
    and goods for each category, the reduction is polynomial in the number of
    agents, goods and categories.
\end{proof}

\section{Reduced Instances}\label{sec:reduced-instances}

High-valued goods are generally harder to handle than low-valued goods in
MMS-approximation. Low-valued goods can easily be distributed across bundles in
an approximately even manner and to a certain extent in a way that makes up for
an uneven value distribution due to the high-valued goods. High-valued goods, on
the other hand, allow only for a rough and usually uneven distribution. In order
to simplify the problem instances, we wish to minimize both the number of
high-valued goods and the maximum value of a good.

If we remove an agent $i$ and a bundle $B\subseteq M$ from an instance, the
result is called a \emph{reduced} instance. If the bundle's value is
sufficiently high ($v_i(B)\geq\alpha\mu_i$) and the MMS of the remaining agents
are at least as high after the removal, this is called a \emph{valid
reduction}~\cite{garg_improved_2020}, a concept used in many MMS approximation
algorithms for the unconstrained fair allocation problem
\citep[e.g.,][]{garg_approximating_2019,kurokawa_fair_2018,ghodsi_fair_2018}.\footnote{The
term \emph{reduction} here refers to data reduction, as the term is used in
parameterized algorithm design, rather than to the problem transformations of
complexity theory.}
With a valid reduction we can both guarantee agent $i$ a bundle with a value of
at least $\alpha\mu_i$ and reduce the original instance to a smaller problem
instance.

Given the above definition, a valid reduction could leave an instance without
any feasible (complete) allocations, as there may be more goods left in a
category than can be allocated to the remaining agents.
We require that a valid reduction leaves the reduced instance with at least one
feasible allocation.

\begin{definition}\label{def:general-reduction}
    Let $I = \langle N, M, V, C \rangle$ be an instance of the fair allocation
    problem under cardinality constraints, $B$ a feasible bundle, $i$ an agent,
    and $I' = \langle N \setminus \{i\}, M \setminus B, V', C' \rangle$, where
    $V'$ and $C'$ are equivalent to $V$ and $C$, with agent $i$ and the items in
    $B$ removed. If $v_i(B) \ge \alpha\mu_i^I$, $\mathcal{F}_{I'} \neq
    \emptyset$ and $\mu_{i'}^{I'} \ge \mu_{i'}^{I}$ for all $i' \in N \setminus
    \{i\}$, then allocating $B$ to $i$ is called a \textit{valid reduction}.
\end{definition}

\noindent Most of the valid reductions used in unconstrained fair allocation are
based on the pigeonhole principle. If you can find a set of goods that are worth
at least $\alpha\mu_i$ to some agent $i$ and show that all agents must have an
MMS partition with a bundle containing an equivalent number of equally or higher
valued goods, then you have a valid reduction. The latter part is exactly what
the pigeonhole principle promises if we, e.g., look at the bundle $\{n, n + 1\}$
in unconstrained fair allocation. Under cardinality constraints, we can also
utilize the pigeonhole principle to find valid reductions. The usefulness is,
somewhat reduced, due to both a lack of a common preference ordering
across categories and the restrictiveness of the category thresholds. We can,
however, show a general result for valid reductions based on the pigeonhole
principle.

\begin{theorem}\label{thr:general-reduction}
    Let $I = \langle N, M, V, C\rangle$ be an ordered instance of the fair
    allocation problem under cardinality constraints, and let $B = \{j_1, \dots,
    j_k\}$ be a feasible bundle of $k \ge 1$ goods such that $v_i(B) \ge
    \alpha\mu_i$ for an agent $i \in N$ and $\alpha > 0$. Let each
    agent $i' \in N \setminus \{i\}$ have a bundle $B_{i'}$ in one of her MMS
    partitions such that there is an injective map $f : B \rightarrow B_{i'}$
    where, for each $j \in B$, $j$ and $f(j)$ belong to the same category, and
    $v_{i'}(f(j)) \ge v_{i'}(j)$. Let $B'$ be the bundle consisting of
    the goods in $B$ and for each $C_h \in C$ the $\max(0, |C_h\setminus B| -
    (|N| - 1)k_h)$ lowest-valued goods in $C_h \setminus B$. Then, $B'$ and
    $i$ form a valid reduction for $I$ and $\alpha$.
\end{theorem}

\begin{proof-sketch}
    For any agent $i' \neq i$, the injective map and the construction of $B'$
    guarantees that there is a way to modify the MMS partition of $i'$ through
    trades and transfers of goods, such that one bundle is turned into $B'$ and
    the value of any other bundle is at least as high as in the MMS partition
    originally. The construction of $B'$ also guarantees a valid instance after
    the reduction. Since $v_i(B') \ge v_i(B) \ge \alpha\mu_i$, $B'$ and $i$ form
    a valid reduction for $I$ and $\alpha$.
    \qed
\end{proof-sketch}

\noindent We can easily use the general result of \cref{thr:general-reduction}
to construct similar valid reductions to those in the unconstrained setting.
Any good $i$ valued at more than $\alpha\mu_i$ for some agent $i$ can be used
for a reduction, as the identity function $f : \{j\} \rightarrow \{j\}$
satisfies the criteria of \cref{thr:general-reduction}. Similarly, by the
pigeonhole principle, we can create valid reductions with the $n$-th and $(n +
1)$-th most valuable goods in a single category.

\begin{corollary}\label{cor:single-item-reduction}
    Let $I = \langle N, M, V, C\rangle$ be an ordered instance of the fair
    allocation problem under cardinality constraints, where there is an agent $i
    \in N$ and a good $j \in M$ such that $v_{ij} \ge \alpha\mu_i$ for $\alpha >
    0$. Then, a valid reduction can be constructed from the bundle $B = \{j\}$.
\end{corollary}

\begin{corollary}\label{cor:double-item-reduction}
    Let $I = \langle N, M, V, C\rangle$ be an ordered instance of the fair
    allocation problem under cardinality constraints, with a category $C_h =
    \langle c_1, c_2, \dots, c_{|C_h|} \rangle$, $|C_h| \ge |N| + 1$, where
    $v_{i}(\{c_{|N|}, c_{|N| + 1}) \ge \alpha\mu_i$ for some $i \in N$
    and $\alpha > 0$.  Then, a valid reduction can be constructed from the
    bundle $B = \{c_{|N|}, c_{|N| + 1}\}$.
\end{corollary}

\noindent It can be tempting to think that we can employ the same valid
reductions within a single category as is possible in the unconstrained setting.
This is not the case, even when the instance only has a single category and
three agents with identical valuations. For example, in the unconstrained
setting, any bundle $B$ consisting of two goods, with $v_{i}(B) \ge \alpha\mu_i$
for an agent $i \in N$ and $v_{i'}(B) \le \mu_{i'}$ for all other agents $i' \in
N \setminus \{i\}$, can be used for a valid reduction. This, is not the case
under cardinality constraints, even when removing $B$ and $i$ produces a
feasible instance without removing any other goods.\footnote{See
\cref{exm:failing-reduction} in the appendix for a simple instance where this
fails.}

\section{MMS Results under Cardinality Constraints}\label{sec:approx-alg}
\label{sec:general-results}

The reductions of \cref{thr:normalized,thr:ordered,cor:single-item-reduction},
which can be performed in polynomial-time, let us restrict
finding $\alpha$-approximate MMS allocations to normalized ordered instances
where each good is worth less than $\alpha$, without loss of generality. For
such instances, \cref{alg:approx} can be used to find $(|N|/(2|N| -
1))$-approximate MMS allocations, which for any number of agents is at least a
$1/2$-approximate MMS allocation.

The algorithm works in a somewhat similar manner to bag filling algorithms for
unconstrained fair allocation \citep[see,
e.g.,][]{garg_approximating_2019,ghodsi_fair_2018}, i.e., by incrementally
adding goods to (and, in our case, removing goods from) a ``bag,'' or partial
bundle, $B$, until $v_i(B) \ge \alpha$ for some agent $i$. The major difference
is the initial content of the bundle. To make sure that a complete feasible
allocation is found, the bundle initially contains the $\lfloor |C_h|/n \rfloor$
least-valuable remaining goods in each category $C_h$ (denoted by $C_h^L$). This
guarantees that the required number of goods is given away from each category.
The value of the bundle is then incrementally increased, so as to not increase
the value by more than $\alpha$ in each step, by exchanging one of the goods in
$B$ from some $C_h^L$, for one of the $\lfloor |C_h|/n \rfloor$ most valuable
remaining goods in the same category (denoted $C_h^H$). To mitigate possible
effects of rounding $|C_h|/n$, one additional good may be added from any
category where $|C_h| / n > \lfloor |C_h|/n \rfloor$.

\begin{figure}[t]
\begin{procedure}{Find a $\alpha$-MMS solution to ordered instance}{approx}
\Preamble{A normalized ordered instance $I = \langle N, M, V, C \rangle$ with
all $v_{ij} < \alpha$}{Allocation $A$ consisting of each bundle $B$ allocated}
\begin{pseudo}
    \kw{while} there is more than one agent left \\+
        $B$ = $\cup_{h = 1}^{\ell} C_h^L$ \\
        \kw{while} $v_i(B) < \alpha$ for all agents $i$ \label{ln:whilelowval} \\+
            \kw{if} $B \cap C_h^L \neq \emptyset$ for some $C_h$ \\+
                $j =$ any element of $C_h^H \setminus B$ \\
                $j' =$ any element of $B \cap C_h^L$ \\
                $B = (B \setminus \{j'\})\cup \{j\}$ \\-
            \kw{else} $j =$ any $c_{\lceil |C_h|/n \rceil}$ not in $B$ \\+
                $B = B \cup \{j\}$ \\--
        allocate $B$ to some agent $i$ with $v_i(B) \ge \alpha$ \label{ln:bundle}\\
        remove $B$ and $i$ from $I$ and update $n$, and $C_h^H$ and $C_h^L$ for
        all $h$ \\-
        allocate the remaining goods to the last agent
\end{pseudo}
\end{procedure}
\end{figure}

Before proving that the algorithm does indeed find a $1/2$-approximate MMS
allocation, we first need a lower bound on the value of the remaining goods at
any point during the execution of the algorithm.

\begin{lemma}\label{lem:remaining-value}
    Let $I = \langle N, M, V, C \rangle$ be a normalized ordered instance of the
    fair allocation problem under cardinality constraints where all goods are
    worth less than $\alpha$ for some $\alpha \ge 1/2$. Let $n$ denote the
    number of remaining agents at any given point during the execution of
    \cref{alg:approx}.  Then each remaining agent assigns a value of at least
    $|N| - 2(|N| - n)\alpha$ to the set of unallocated goods.
\end{lemma}

\begin{proof}
    Because the instance is normalized, the lemma holds at the start of the
    algorithm. Assume that there are $n$ remaining agents at the start of an
    iteration, and for each remaining agent $i$, $v_i(M) \ge |N| - 2(|N| -
    n)\alpha$. Let $i'$ be the agent receiving $B$ in the iteration. For any
    remaining agent $i \neq i'$, we wish to show that $v_i(M \setminus B) \ge
    |N| - 2(|N| - n +
    1)\alpha$. Because the valuations are additive, the only way
    this cannot hold
    is if $v_i(B) > 2\alpha$. Since any change
    to $B$ after the initial creation adds a good to $B$ or exchanges a good in
    $B$ for another, any individual change cannot increase the value of $B$ by
    more than $\alpha$. Thus, because the loop at line~\ref{ln:whilelowval}
    terminates as soon as $v_i(B)\geq\alpha$, the only way we may have $v_i(B) >
    2\alpha$ is if it holds initially, i.e., $B = \smash{\bigcup_{h = 1}^\ell}
    C_h^L$ and $v_i(\smash{\bigcup_{h = 1}^\ell} C_h^L) > 2\alpha$. However, by
    definition $v_i(C_h^L) \le v_i(C_h)/n$ which implies $v_i(B) \le v_i(M)/n$.
    Consequently, $v_i(M \setminus B) \ge (n - 1)v_i(B) \ge (n - 1)2\alpha \ge
    (n - 1) \ge |N| - 2(|N| - n + 1)\alpha$.
    \qed
\end{proof}

\noindent With \cref{lem:remaining-value} we have a sufficient lower guarantee
for the remaining value. We are now ready to show the guarantees of the
algorithm.

\begin{lemma}\label{lem:correctness}
    Given a normalized ordered instance $I = \langle N, M, V, C \rangle$ of the
    fair allocation problem under cardinality constraints where all goods are
    worth less than $\alpha = |N|/(2|N| - 1)$, \cref{alg:approx} finds a
    feasible $(|N|/(2|N| - 1))$-approximate MMS allocation in polynomial time
    in the number of agents and goods.
\end{lemma}

\begin{proof}
    When allocating the remaining goods to the last agent,
    \cref{lem:remaining-value} guarantees that the goods are worth at least
    $\alpha$, if $|N| - 2(|N| - 1)\alpha \ge \alpha$, which holds for $\alpha
    \le |N|/(2|N| - 1)$. Additionally, as long as $B$ reaches a value of
    $\alpha$ before running out of improvement operations, any other agent is
    also guaranteed to receive a bundle they value at no less than $\alpha$.
    Since $B$ contains the $\lceil C_h/n \rceil$ most valuable goods in each
    category $C_h$ when the algorithm runs out of operations, $B$ reaches a
    value of at least $1/n$ of the remaining value. We thus only need to show
    that the remaining value is always at least $n\alpha$ for any remaining
    agent. \Cref{lem:remaining-value} guarantees that the remaining value is at
    least $|N| - 2(|N| - n)\alpha$. Since, this value is at least $\alpha$ for
    $n = |N| - 1$, the value is at least $2(n - 1)\alpha + \alpha \ge n\alpha$
    for any other $n$, and we are guaranteed that the value of $B$ reaches at
    least $\alpha$ in any iteration. Since $\mu_i \le 1$ for $i \in N$, each
    agent $i$ receives at least $\alpha\mu_i$ value.

    It remains to show that any bundle allocated is feasible. As long as $|C_h|
    \le nk_h$, it holds that $\lceil |C_h|/n \rceil \le k_h$ and any bundle
    allocated is feasible. Obviously, $|C_h| \le nk_h$ holds when $n = |N|$, as
    all instances are assumed to have at least one feasible complete allocation.
    Assume that $|C_h| \le nk_h$ holds at the start of an iteration. The bundle
    $B$ starts with $\lfloor |C_h|/n \rfloor \ge |C_h| - (n - 1)k_h$ of the
    goods in $C_h$ and no good is removed without adding another from the same
    $C_h$. Thus, $|C_h \setminus B| \le (n - 1)k_h$ and the condition holds for
    $n - 1$ after allocating $B$. Consequently, each allocated bundle, including
    the bundle allocated to the last agent, is feasible.

    In each iteration of the algorithm, goods are added to and exchanged through
    a set of operations. As no good is added back into $B$ after being
    removed, the number of operations in each iteration is polynomial in the
    number of agents and goods. Since there are $|N| - 1$ iterations, the
    running time of the algorithm is also polynomial in the number of agents and
    goods.
    \qed
\end{proof}

\noindent We have now showed everything needed to show that $1/2$-approximate
MMS allocations exist and can be found in polynomial time.

\begin{theorem}\label{thr:correctness}
    For an instance $I = \langle N, M, V, C \rangle$ of the fair allocation
    problem under cardinality constraints, a $(|N|/(2|N| - 1))$-approximate MMS
    allocation always exists and can be found in polynomial time.
\end{theorem}

\begin{proof}
    By \cref{thr:normalized,thr:ordered,cor:single-item-reduction}, any instance
    $I$ can in polynomial time be converted to one, $I'$, that \cref{alg:approx}
    accepts. Since $I'$ has no more agents than $I$, \cref{lem:correctness}
    guarantees that for $I'$ an at least $(|N|/(2|N| - 1))$-approximate MMS
    allocation is found in polynomial time by \cref{alg:approx}. The allocation
    for $I'$ can then be turned back to one for $I$ in polynomial time.
    \qed
\end{proof}

\noindent \Cref{alg:approx} is guaranteed to find $\alpha$-approximate MMS
allocations for \emph{all} possible problem instances when $\alpha \le |N|/(2|N|
- 1)$. However, there exist many types of problem instances for which the
algorithm will find a feasible $\alpha$-approximate MMS allocation when a larger
$\alpha$ is used. For example, for an instance where $v_{ij} \le \mu_i/4$ for
all $i \in N$, $j \in M$, the algorithm will always find a feasible
$\alpha$-approximate MMS allocation when $\alpha = 3/4$, because then each
bundle allocated in the bag filling step is worth no more than $1$, unless the
bundle is the starting bag. Generally, increasing $\alpha$ might in the worst
case result in the remaining value decreasing to the point where $v_i(B) <
\alpha$ for any remaining agent $i$ after all improvements have been performed
on $B$. However, for many problem instances, the average value of each allocated
bundle is quite a bit smaller than $2\alpha$ for any remaining agent $i$. Thus,
even for larger values of $\alpha$, the algorithm can often find a
$\alpha$-approximate MMS allocation. While it is hard to determine the largest
$\alpha$ that works for a certain problem instance through calculation, it is
possible to simply check if the algorithm finishes for various values of
$\alpha$. Preliminary experiments suggest that trying the algorithm for a
limited number of different values of $\alpha$ often provides much better
approximations.

Since \cref{thr:correctness} in fact guarantees each agent a bundle of value at
least $(|N|/(2|N| - 1))v_i(M)$, it directly allows us to show that a
$1$-out-of-$(2|N|-1)$ MMS allocation always exists and can be found in
polynomial time.

\begin{corollary}
    For an instance $I = \langle N, M, V, C \rangle$ of the fair allocation
    problem under cardinality constrains, a $1$-out-of-$(2|N| - 1)$ MMS
    allocation always exists and can be found in polynomial time.
\end{corollary}

\begin{proof}
    In a similar fashion to \cref{thr:normalized}, the $1$-out-of-$(2|N| - 1)$
    MMS of any agent can at most be $v_i(M)/(2|N| - 1)$. The proof of
    \cref{lem:correctness} shows that \cref{alg:approx} gives each agent a
    bundle valued at least $|N|/(2|N| - 1)$ when $v_i(M) = |N|$, which is at
    least the $1$-out-of-$(2|N| - 1)$ MMS of any agent.
    \qed
\end{proof}

\noindent It is possible to improve the existence guarantee for MMS
approximation by using bag filling in combination with the lone-divider
technique of \citeauthor{Aigner-Horev:2022} \cite{Aigner-Horev:2022}. In the
lone-divider technique, agent $i$, one of the remaining agents, is chosen to
partition the remaining goods into bundles that all have a value of at least
$\alpha\mu_i$ to $i$. Then, a non-empty subset of the bundles is allocated to
some subset of the remaining agents, through an \emph{envy-free matching} which
is guaranteed to exist. An envy-free matching is here a matching where each
agent matched to a bundle values it at no less than $\alpha\mu_i$ and all
non-matched remaining agents value the matched bundles at less than
$\alpha\mu_i$. \citeauthor{Aigner-Horev:2022} showed that an envy-free matching
always exists \cite{Aigner-Horev:2022}. The process is then repeated until no
agent remains. In order to improve the existence guarantee, we first use the
lone-divider technique with a partition scheme that only works when a large
number of agents remain. When the partition scheme no longer works, the ratio of
remaining value to remaining agents has increased, since $\alpha < 1$ and any
bundle already allocated is worth less than $\alpha\mu_i$ to any remaining
agent. The increased ratio allows \cref{alg:approx} to be able to provide each
remaining agent with a greater value than before the allocations. Unfortunately,
the existence result is only of an existential nature, as the partition scheme
depends on finding arbitrary MMS-partitions, which is known to be NP-hard
\cite{woeginger_polynomial-time_1997}.

\begin{theorem}\label{thr:MMS-existence-improvement}
    For an instance $I = \langle N, M, V, C \rangle$ of the fair allocation
    problem under cardinality constraints, a $(\sqrt{|N|}/(2\sqrt{|N|} -
    1))$-approximate MMS allocation always exists.
\end{theorem}

\begin{proof-sketch}
    When only a few bundles have been given away, any MMS-partition of $I$ for
    any remaining agent contains at least as many bundles with a remaining value
    of $\alpha\mu_i$ or higher, as there are remaining agents. The goods in the
    other bundles in the MMS-partition can then arbitrarily be moved to one of
    these bundles with remaining value $\alpha\mu_i$. On the other hand, since
    $\alpha < 1$, as the number of allocated bundles increases, each remaining
    agent's proportional share of the value of the remaining goods increases.
    Thus, \cref{alg:approx} will be able to guarantee a partition with higher
    and higher minimum bundle value. The value of $\alpha$ must then be set so
    that in any situation, one of the two methods works. It can be shown that
    $\sqrt{|N|}/(2\sqrt{|N|} - 1)$ is the largest value of $\alpha$ that
    works.
    \qed
\end{proof-sketch}

\section{Uniform Matroid Constraints}\label{sec:uniform-matroid-constraints}

In this section we deal with the special case of cardinality constraints in
which there is only a single category, i.e., $\ell = 1$. In this case, the
cardinality constraints are equivalent to simply limiting the maximum number of
goods in a bundle, or, equivalently, restricting bundles to be independent sets
of a \emph{uniform matroid}. Throughout the section we will assume that for any
ordered instance, which provides a total ordering of goods, the goods are
numbered in a way such that $v_i(j) \ge v_i(j')$ for all $i \in N$ and $j, j'
\in M$ with $j < j'$. In other words, the goods are numbered from most preferred
(1) to least preferred ($|M|$). Our main result
(\cref{thr:mms-2/3-single-category}) for single-category instances is the
existence of $(2/3)$-approximate MMS allocations and the ability to find these
in polynomial time.

\begin{theorem}\label{thr:mms-2/3-single-category}
    For an instance $I = \langle N, M, V, \langle (C_1, k_1) \rangle \rangle$ of
    the fair allocation problem under cardinality constraints, a
    $(2/3)$-approximate MMS allocation always exists and can be found in
    polynomial time.
\end{theorem}

\noindent
In order to prove \cref{thr:mms-2/3-single-category} we need the following
observation about the value of certain subsets of goods.

\begin{lemma}\label{lem:single-category-value-adjustment}
    Let $I = \langle N, M, V, \langle (C_1, k_1) \rangle \rangle$ be an ordered
    instance of the fair allocation problem under cardinality constraints. For
    any $r \in \{1, 2, \dots, |N|\}$, let $B_r = \{r, r + 1, \dots,
    \min(|M|, r + k_1(|N| - r + 1) - 1)\}$. Then, for any $i \in N$,
    \(
        v_i(B_r) \ge (|N| - r + 1)\mu_i
    \).
\end{lemma}

\noindent
\Cref{lem:single-category-value-adjustment} provides two useful properties.
Most importantly, it can be used to show that the bundles created during a
bag-filling style algorithm (\cref{alg:2/3-MMS}) will be worth at least $\mu_i$
before running out of improvements. At the same time, it provides a direct,
polynomial way to improve our estimate of $\mu_i$ (in addition to
\cref{thr:normalized}) to the required accuracy for the algorithm.
\Cref{lem:single-category-value-adjustment} can be used to show that $2/3$-MMS
allocations can be found in polynomial time for a restricted class of instances
using \cref{alg:2/3-MMS}.

\begin{lemma}\label{lem:alg-2/3-MMS-correctness}
    For an instance $I$ of the fair allocation problem under cardinality
    constraints satisfying the requirements of \cref{alg:2/3-MMS},
    the algorithm finds a $2/3$-approximate MMS allocation in polynomial time.
\end{lemma}

\begin{proof-sketch}
    The correctness of \cref{alg:2/3-MMS} follows from two observations about
    the construction of $B'_j$. First, the construction guarantees that
    $B'_j$ is feasible and contains at least the required number of goods so
    that after allocating $B'_j$, there are at most $k_1(j - 1)$ goods left.
    Second, \cref{lem:single-category-value-adjustment}, the incremental
    improvements of $B'_j$ and the distribution of the $|N|$ most valuable goods
    into distinct bundles, together guarantee that when $j = r$, the value of
    the $\min(k_1, |B_r \cap M|)$ most valuable remaining goods in $B_r$ is at
    least $1$ for each remaining agent. Thus, $B'_j$ will always be able to
    reach a value of at least $2/3$.
    \qed
\end{proof-sketch}

\begin{figure}[t]
    \begin{procedure}{Find $(2/3)$-MMS solution for single-category
        instance}{2/3-MMS}
        \Preamble{An ordered instance $I = \langle N, M, V, \langle C_1, k_1
        \rangle \rangle$ with $|M| > |N|$, $\mu_i \le 1$, $v_i(B_r) \ge
        |N| - r + 1$ (from \cref{lem:single-category-value-adjustment}),
        $v_i(1) < 2/3$, and $v_i(|N| + 1) < 1/3$ for every $i \in N$, $r
        \in \{1, 2, \dots, |N|\}$}{Allocation $A$ consisting of each bundle
        $B'_j$ allocated}
        \begin{pseudo}
            let $B'_1 = \{1\}, B'_2 = \{2\}, \dots, B'_{|N|} = \{|N|\}$ \\
            \kw{for} $j = |N|$ down to $1$ \\+
                \kw{if} $|M| > k_1(j - 1) + 1$ \\+
                    add the $|M| - k_1(j - 1) - 1$ least-valuable goods in
                    \label{line:2/3-MMS-minimum-goods}\\+*
                  & $M \setminus (B'_1 \cup B'_2 \cup \dots \cup B'_j)$ to
                    $B'_j$ \\--
                \kw{while} $v_i(B'_j) \le 2/3$ for all $i \in N$ and $|B'_j| <
                k_1$ \label{line:2/3-MMS-item-adding}\\+
                    add the least-valuable good in $M \setminus (B'_1 \cup
                    B'_2 \cup \dots \cup B'_j)$ to $B'_j$ \\-
                \kw{while} $v_i(B'_j) \le 2/3$ for all $i \in N$
                \label{line:2/3-MMS-item-upgrade}\\+
                    exchange the least valuable $g \in B'_j$ for the least
                    valuable \\+*
                  & $g' \in M \setminus (B'_1 \cup B'_2 \cup \dots \cup
                    B'_j)$ with $g' < g$ \\--
                find $i \in N$ such that $v_i(B'_j) \ge 2/3$ \\
                allocate $B'_j$ to $i$ and set $N = N \setminus \{i\}$, $M
                = M \setminus B'_j$.
        \end{pseudo}
    \end{procedure}
\end{figure}

\begin{proof-sketch}[for \cref{thr:mms-2/3-single-category}]
    The proof boils down to showing that for any instance $I$, we can either
    trivially, if $|M| \le |N|$, find a $(2/3)$-approximate MMS allocation
    through valid reduction, or we can turn $I$ into an instance accepted by
    \cref{alg:2/3-MMS}. The latter is achieved through repeated rescaling based
    on \cref{thr:normalized} and \cref{lem:single-category-value-adjustment},
    together with applying all possible valid reductions based on
    \cref{cor:single-item-reduction,cor:double-item-reduction}.
    \qed
\end{proof-sketch}

\noindent In addition to existence of
$(2/3)$-approximate MMS allocations, certain restricted classes of
single-category instances allow for better approximation or existence
guarantees. Specifically, when the number of goods is not much larger than the
category threshold, approximation results for unconstrained fair allocation
apply under cardinality constraints.

\begin{lemma}\label{lem:few-items-reduction-to-unconstrained}
    For an instance $I = \langle N, M, V, \langle (C_1, k_1) \rangle \rangle$ of
    the fair allocation problem under cardinality constraints, with $|M| < |N| +
    k_1$, MMS-approximation reduces to MMS-approximation for unconstrained fair
    allocation.
\end{lemma}

\noindent As a result of \cref{lem:few-items-reduction-to-unconstrained}, the
following follows directly from the results of \citeauthor{garg_improved_2020}
on MMS approximation in unconstrained fair allocation \cite{garg_improved_2020}.

\begin{corollary}
    For an instance $I = \langle N, M, V, \langle (C_1, k_1)\rangle \rangle$ of
    the fair allocation problem under cardinality constraints, with $|M| < |N| +
    k_1$, a $(3/4 + 1/(12n))$-approximate MMS allocation always exists and
    a $(3/4)$-approximate MMS allocation can be found in polynomial time.
\end{corollary}

\noindent When the threshold is small enough, it is possible to show that MMS
allocations always exist. For larger thresholds, on the other hand, it is
possible to create instances for which there is no MMS allocation.

\begin{lemma}\label{lem:single-category-low-threshold}
    Let $I = \langle N, M, V, \langle (C_1, k_1)\rangle \rangle$ be an instance
    of the fair allocation problem under cardinality constraints. If $k_1 \le
    2$, an MMS allocation always exists. If $k_1 \ge 4$, an MMS allocation is
    not guaranteed to exist.
\end{lemma}

\section{Fair Allocation of Chores}\label{sec:chores}

So far we have only considered instances where the items are goods. In this
section we instead consider instances where the items are chores. As our results
on chores are similar in scope and technique to our results on goods, the
results will only be covered briefly with all proofs
given in the appendix. We assume, without loss
of generality, that $v_i(M) < 0$.\footnote{As with goods, normalization does not
work if $v_i(M) = 0$. In this case, $i$ can be removed from the (ordered)
instance by allocating $i$ the $k_h$ worst chores in each $C_h$. This would
constitute a valid reduction.} Then concepts of scale invariance and
normalization transfer directly to chores.

\begin{theorem}[Scale invariance]\label{thr:scale-invariance-chores}
    If $A$ is an MMS allocation for the instance $I = \langle N, M, V, C
    \rangle$ of the fair allocation of chores problem under cardinality
    constraints, then $A$ is also an MMS allocation of $I' = \langle N, M, V', C
    \rangle$, where $v'_i(S) = a_iv_i(S)$, $a_i > 0$, for some agent $i$.
\end{theorem}

\begin{theorem}[Normalization]\label{thr:normalization-chores}
    Let $I = \langle N, M, V, C \rangle$ be an instance of the fair allocation
    of chores problem under cardinality constraints and $v_i(M) = -|N|$ for some
    agent $i$. Then $\mu_i \le -1$.
\end{theorem}

\noindent Further, the reduction to ordered instances works for chores as well.
As with goods, reassigning the valuations of the chores does not change the MMS
of any agent.
The
earlier conversion algorithm for an allocation
of the ordered instance provides each agent with a bundle of equal or higher
value (less disutility), which provides an equal or better approximation.

\begin{theorem}\label{thr:ordered-instance-reduction-chores}
    For fair allocation of chores under cardinality constraints,
    MMS-approximation reduces to MMS-approximation of ordered instances in
    polynomial time.
\end{theorem}

\noindent For chores, the use of valid reductions does not make sense in the
same way as for goods. While valid reductions could still exist and be
used,
there is a lack of simple rules for finding
useful valid reductions. However, we can still bound the number of chores that
have a
large disutility by exploiting the pigeonhole principle on MMS partitions.
Note that \cref{thr:value-adjustment-chores} provides a stronger upper bound on
the number of high-valued chores than the bounds for goods when $\ell \ge 2$.

\begin{theorem}\label{thr:value-adjustment-chores}
    Let $I = \langle N, M, V, C \rangle$ be an instance of the fair allocation
    of chores problem under cardinality constraints, with $|M| \ge |N|r + 1$ for
    an $r \in \{0, 1, \dots\}$. For agent $i \in N$, let $g_{i_j} \in M$ denote
    the $j$-th most valuable chore in $M$ for $i$. Then,
    \[
        v_i(\{g_{i_{|N|r + 1 - r}}, g_{i_{|N|r + 2 - r}}, \dots,
        g_{i_{|N|r + 1}}\}) \ge \mu_i
    \]
\end{theorem}

\noindent
\Cref{thr:scale-invariance-chores,thr:normalization-chores,thr:value-adjustment-chores}
allow for an easy adjustment of the valuation functions such that for each agent
$i \in N$, $\mu_i \le -1$, $v_i(M) \ge -|N|$ and there are at most $r|N|$
chores that $i$ values at less than $-1/(r + 1)$. Crucially, this guarantees
that no chore is valued at less than $-1$, allowing a variant of the bag-filling
algorithm used for goods to find $2$-approximate MMS allocations.

\begin{theorem}\label{thr:2-MMS-chores}
    For an instance $I = \langle N, M, V, C \rangle$ of the fair allocation of
    chores problem under cardinality constraints, a $((2|N| -
    1)/|N|)$-approximate MMS allocation always exists and can be found in
    polynomial time.
\end{theorem}

\noindent For single-category instances we can also for chores find much better
MMS approximate allocations using an algorithm similar to \cref{alg:2/3-MMS}.

\begin{theorem}\label{thr:3/2-MMS-chores-single-category}
    For an instance $I = \langle N, M, V, \langle (C_1, k_1) \rangle \rangle$ of
    the fair allocation of chores problem under cardinality constraints, a
    $(3/2)$-approximate MMS allocation always exists and can be found in
    polynomial time.
\end{theorem}

\section{Discussion}

We improved the currently best known MMS approximation guarantees for
cardinality constraints by extending the concepts of ordered instances and valid
reductions to this setting. Cardinality constraints do, however, impose
additional challenges that do not exist in the unconstrained setting, limiting
the achievable approximation guarantees. The apparent lack of a common
preference ordering between distinct categories limits the degree to which the
number of and maximum value of high-valued goods can be restricted---an
important factor in improving the approximation guarantee of bag-filling style
algorithms. Cardinality constraints also restrict the usability of other types
of MMS-approximation algorithms. For example, the lone-divider method may easily
allocate bundles that contain many items from a single category and few from
others, which in turn can make all further feasible divisions very unbalanced.

\subsubsection{Experiments.} An earlier version of this preprint
(\href{https://arxiv.org/abs/2106.07300v1}{v1}) contains some preliminary
experimental results, along with source code.

\subsubsection{Acknowledgements.} The authors wish to acknowledge valuable input
from anonymous reviewers.

\bibliographystyle{splncs04}
\bibliography{paper}


\newpage
\appendix
\begin{center}
    \huge \textbf{Appendix}
\end{center}

\section{Omitted Proofs for \cref{sec:preliminaries}}

\begin{proof}[for \cref{thm:scale-inv}]\label{proof:thm:scale-inv}
    Assume that $A$ is not an MMS allocation of $I'$. Then, there is an agent
    $i' \in N$ such that $v'_{i'}(A_{i'}) < \mu_{i'}^{I'}$. However, since
    $v'_{i'}(j) = a_{i'}v_{i'}(j)$, it follows that
    \[
        \mu_{i'}^{I'} = \max_{A' \in \mathcal{F}_{I'}}\min_{A_j \in A'}
        v'_{i'}(A_j) = \max_{A' \in \mathcal{F}_{I'}}\min_{A_j \in A'}
        a_{i'}v_{i'}(A_j) = a_{i'}\mu_{i'}^I
    \]

    \noindent Consequently, $v'_{i'}(A_{i'}) = a_{i'}v_{i'}(A_{i'}) \ge
    a_{i'}\mu_{i'}^I = \mu_{i'}^{I'}$. This is a contradiction, and there can be
    no $i'$ with $v'_{i'}(A_{i'}) < \mu_{i'}^{I'}$. Hence, $A$ is an MMS
    allocation of $I'$.
\end{proof}

\begin{proof}[for \cref{thr:normalized}]
    Since the valuations are additive, for any allocation $A$ of $I$, we have
    that $v_i(M) = \sum_{A_j \in A} v_i(A_j)$. Consequently, the value of the
    least valuable bundle $B^* \in A$ must be such that $|N|v_i(B^*) \le v_i(M)$,
    as otherwise, $\sum_{A_j \in A} v_i(A_j) > v_i(M)$. Therefore,
    \[
        \mu_i = \max_{A \in \mathcal{F}_I}\min_{A_j \in A} v_i(A_j) \le \max_{A
        \in \mathcal{F}_I} \frac{v_i(M)}{|N|} = 1
    \]
    \qed
\end{proof}

\section{Omitted Proofs for \cref{sec:ordered}}

\begin{proof}[for \cref{lem:ordered}]
    First, note that creating an ordered instance does not affect any agent's
    MMS:
    For any given agent, \cref{alg:convert} implicitly defines a one-to-one
    mapping on each category $C_h$, corresponding to a permutation of the good
    values. Because the valuations are only interchanged within each $C_h$, the
    map can be used to convert any allocation between the two instances,
    preserving feasibility, without changing the value of any individual bundle.
    The MMS of the agent is independent of the valuations of other agents, and
    so the MMS must be the same in both instances.

    As \cref{alg:convert,alg:recover} are equivalent to those of
    \citeauthor{bouveret_characterizing_2016} within each individual category,
    it follows that the value an agent receives from each category is at least as high in
    allocation $A$ for the original instance as in allocation $A'$ for the
    ordered instance. Thus, the agent's total value in $A$ for the original
    instance must at least be as high as in $A'$ for the ordered instance.
    Neither algorithm will introduce violations of the cardinality constraints,
    and since the MMS is unchanged between the two instances, it follows that
    the new allocation is also a feasible $\alpha$-approximate MMS allocation.
    \qed
\end{proof}

\section{Omitted Proofs from \cref{sec:reduced-instances}}

\begin{proof}[for \cref{thr:general-reduction}]
    For $B'$ and $i$ to be a valid reduction we must show that
    \begin{inl}
        \item\label{req:general-reduction-1} $B'$ is a feasible bundle,
        \item\label{req:general-reduction-2} $v_i(B') \ge \alpha\mu_i$,
        \item\label{req:general-reduction-3} $I' = \langle N \setminus \{i\}, M
            \setminus B', V', C' \rangle$ has at least one feasible allocation
            and
        \item\label{req:general-reduction-4}$ \mu_{i'}^{I'} \ge \mu_{i'}^I$ for
            all $i' \in N \setminus \{i\}$
    \end{inl}.

    For $B'$ to be feasible, then it must hold that $|C_h \cap B'| \le k_h$ for
    any category $C_h \in C$. Since $B$ is feasible ($|C_h \cap B| \le k_h$), we
    have that:
    \begin{align*}
        |C_h \cap B'| &= |C_h \cap B| + \max(0, |C_h \setminus B| - (|N| - 1)k_h)
        \\
            &= |C_h \cap B| + \max(0, |C_h| - |C_h \cap B| - (|N| - 1)k_h) \\
            &\le |C_h \cap B| + |N|k_h - |C_h \cap B| - (|N| - 1)k_h \\
            &\le k_h.
    \end{align*}

    \noindent Consequently, \ref{req:general-reduction-1} holds. Further, since
    $B \subseteq B'$ we have $v_i(B') \ge v_i(B) \ge \alpha\mu_i$ and
    \ref{req:general-reduction-2} also holds. For \ref{req:general-reduction-3},
    note that if $|C_h \setminus B| > (|N| - 1)k_h$ for some $C_h \in C$,
    $B'$ contains $|C_h \setminus B| - (|N|-1)k_h$ additional goods from
    $C_h$. Consequently, $|C_h \setminus B'| \le (|N| - 1)k_h$ and $C_h
    \setminus B'$ does not contain more goods than can be given to $|N| - 1$
    agents. Therefore, $I'$ has a feasible allocation and
    \ref{req:general-reduction-3} holds.

    For simplicity, we now assume without loss of generality that if $j \in B$
    and $j \in B_{i'}$, then $f(j) = j$. If this does not hold for $f$, then we
    can define a new injective function $f' : B \rightarrow B_{i'}$ satisfying
    the criteria of the theorem by
    \[
        f'(j') =
        \begin{cases}
            j & \text{if } j' = j \\
            f(j) & \text{if } f(j') = j \\
            f(j') & \text{otherwise}
        \end{cases}
    \]
    for all $j' \in B$. The requirements of the theorem also hold for $f'$, as
    it is injective, $j$, $j'$ and $f(j)$ all belong to the same category and
    $v_{i'}(j') \le v_{i'}(j) \le v_{i'}(f(j))$.

    For \ref{req:general-reduction-4} we want to show that for any agent $i' \in
    N \setminus \{i\}$, we can modify their MMS partition such that $B_{i'}$ is
    converted into $B$ while maintaining a feasible partition and without
    decreasing the value of any other bundle in the partition. In other words,
    we want to show that there is a feasible partition containing $B$, and
    $(|N|- 1)$ bundles with a value of no less than $\mu_{i'}^I$ to $i'$. The
    transformation can be achieved by performing three steps in order:

    \begin{enumerate}
        \item\label{item:general-reduction-1} For each $j \in B$, exchange the
            placement of $j$ and $f(j)$ in the MMS partition.
        \item\label{item:general-reduction-2} While $|C_h \cap B'| < |C_h \cap
            B_{i'}|$, move a good in $(C_h \cap B_{i'}) \setminus B'$ to any
            other bundle $B_{i''}$ with $|C_h \cap B_{i''}| < k_h$.
        \item\label{item:general-reduction-3} While $C_h \cap B' \neq C_h \cap
            B_{i'}$, exchange any good in $(C_h \cap B_{i'}) \setminus B'$ for a
            good of equivalent or lower value in $(C_h \cap B')\setminus
            B_{i'}$.
    \end{enumerate}

    \noindent Note that non of the operations can cause infeasibility, as both
    goods in steps~\ref{item:general-reduction-1} and
    \ref{item:general-reduction-3} belong to the same category. Additionally, in
    step~\ref{item:general-reduction-2}, $B_{i''}$ has space for at least one
    more good from $C_h$.

    The value of any other bundle than $B_{i'}$ will not decrease during the
    transformation. In steps~\ref{item:general-reduction-1} and
    \ref{item:general-reduction-3} the good removed from $B_{i'}$ has a value
    that is either equivalent to or higher than the one it is exchanged for. As
    each good has a non-negative value, the value of the receiving bundle cannot
    decrease in step~\ref{item:general-reduction-2}. Since each of the bundles
    in the MMS partition had a value of at least $\mu_i^I$ prior to the
    transformation, there are at least $|N| - 1$ bundles with a value of at
    least $\mu_i^I$ after the transformation.

    We now need to show that each step can be performed, and that $B' = B_{i'}$
    after step~\ref{item:general-reduction-3}. Exchanging the position of two
    goods from the same category is always possible, and
    step~\ref{item:general-reduction-1} can always be performed. Additionally,
    as $f$ is injective and $f(j) = j$ for any $j \in B$ with $j \in B_{i'}$,
    we have that for $j \in B$ either $f(j) \notin B$ or $f(j) = j$.
    Consequently, each exchange either brings $j$ into $B_{i'}$ without removing
    any item from $B$ or exchanges $j$ for itself within $B_{i'}$. After
    step~\ref{item:general-reduction-1}, we thus have $B \subseteq B_{i'}$.

    In step~\ref{item:general-reduction-2}, we know that $|(C_h \cap B_{i'})
    \setminus B'| \ge |C_h \cap B_{i'}| - |C_h \cap B'| > 0$ whenever the
    condition of the step holds. The only way the step can fail is thus that
    there is no bundle $B_{i''}$. However, by the construction of $B'$ and as
    $I$ is feasible, we have that when selecting $B_{i''}$:
    \[
        |C_h| \le (|N| - 1)k_h + |C_h \cap B'| \le (|N| - 1)k_h - 1 + |C_h \cap
        B_{i'}|,
    \]
    and one of the other bundles contains less than $k_h$ goods from $C_h$.

    We claim that after step~\ref{item:general-reduction-2}, $|C_h \cap B'| =
    |C_h \cap B_{i'}|$ for each category $C_h \in C$. If this does not hold,
    then $|C_h \cap B'| > |C_h \cap B_{i'}|$ after
    step~\ref{item:general-reduction-1}. We know that either $B \cap C_h = B'
    \cap C_h$ or $|C_h \setminus B'| = (|N| - 1)k_h$.  If $B \cap C_h = B' \cap
    C_h$, then $(B' \cap C_h) \subseteq B_{i'}$ and if $|C_h \setminus B'| =
    (|N| - 1)k_h$ then, $|B' \cap C_h| > |B_{i'} \cap C_h|$ would imply
    infeasiblity. Consequently, $|B' \cap C_h| = |B_{i'} \cap C_h|$ for each
    category $C_h \in C$ after step~\ref{item:general-reduction-2}.
    Additionally, no good in $B'$ is removed from $B_{i'}$ in this step and it
    still holds that $B \subseteq B_{i'}$.

    The only way that step~\ref{item:general-reduction-3} can fail is if $(C_h
    \cap B_{i'}) \setminus B' \neq \emptyset$, but there is no good in $(C_h
    \cap B') \setminus B_{i'}$ of equivalent or lower value. Note that $(C_h
    \cap B') \setminus B_{i'}$ always contains the same number of goods as
    $(C_h \cap B_{i'}) \setminus B'$ as the sets had equivalent size after
    step~\ref{item:general-reduction-2} and each exchange in
    step~\ref{item:general-reduction-3} reduces the size of both by~1. Since $B
    \subseteq B_{i'}$, for any good $j \in (C_h \cap B') \setminus B_{i'}$ we
    have $j \notin B$ and by construction $B'$ contains all goods in $C_h$ of
    lower value than $j$. Therefore, $(C_h \cap B_{i'}) \setminus B'$ cannot
    contain a good of lower value than $j$ and any good in $(C_h \cap B')
    \setminus B_{i'}$ can be used in the exchange. In other words, step 3 will
    never fail.  Finally, after step~\ref{item:general-reduction-3} we have that
    $(C_h \cap B') \setminus B_{i'} = \emptyset = (C_h \cap B_{i'}) \setminus
    B'$ for each category $C_h \in C$ which implies $B' = B_{i'}$. Thus,
    \ref{req:general-reduction-4} holds.
    \qed
\end{proof}

\begin{proof}[for \cref{cor:single-item-reduction}]
    Each good must appear in exactly one bundle in any MMS partition. Thus, each
    agent $i' \in N \setminus \{i\}$ has in each one of their MMS partitions a
    bundle $B_{i'}$, with $j \in B_{i'}$. Consequently, there exists an
    injective map $f : \{j\} \rightarrow B_{i'}$ satisfying the conditions of
    \cref{thr:general-reduction}, namely $f(j) = j$. Since $\{j\}$ is a feasible
    bundle (otherwise $\mathcal{F}_I = \emptyset$), all the conditions of
    \cref{thr:general-reduction} are satisfied and $\{j\}$ can be used to create
    a valid reduction by setting $B = \{j\}$ and creating $B'$ as in
    \cref{thr:general-reduction}.
    \qed
\end{proof}

\begin{proof}[for \cref{cor:double-item-reduction}]
    Since there are exactly $|N|$ bundles in any MMS partition and $|C_h| \ge
    |N| + 1$, any agent $i' \in N \setminus \{i\}$ must by the pigeonhole
    principle in any one of their MMS partitions have a bundle $B_{i'}$ that
    contains at least two goods from $\{c_1, c_2, \dots, c_{|N| + 1}\}$. Since
    $c_{|N|}$ and $c_{|N| + 1}$ are the two least valuable goods in $\{c_1, c_2,
    \dots, c_{|N| + 1}\}$, $B_{i'}$ contains goods $j, j' \in \{c_1, c_2, \dots,
    c_{|N| + 1}\}$, $j \neq j'$, such that $v_{i'}(j) \ge v_{i'}(c_{|N|})$ and
    $v_{i'}(j') \ge v_{i'}(c_{|N| + 1})$. Consequently, the map $f : \{c_{|N|},
    c_{|N| + 1}\} \rightarrow B_{i'}$ with $f(c_{|N|}) = j$ and $f(c_{|N| + 1})
    = j'$ satisfies the conditions of \cref{thr:general-reduction}. Since
    $\mathcal{F_I} \neq \emptyset$, the bundle $\{c_{|N|}, c_{|N| + 1}\}$ is
    feasible and \cref{thr:general-reduction} says that $\{c_{|N|}, c_{|N| +
    1}\}$ can be used to create a valid reduction.
    \qed
\end{proof}

\section{Omitted Proofs from \cref{sec:general-results}}

\begin{proof}[for \cref{thr:MMS-existence-improvement}]
    To show that a certain approximation guarantee can be fulfilled by the
    lone-divider technique, it suffice to show that after $b < |N|$ bundles have
    been allocated, any one of the remaining agents $i$ can partition the
    remaining goods into $|N| - b$ feasible bundles each with a value of at
    least $\alpha\mu_i$. We wish to show that this is possible when $\alpha =
    (\sqrt{|N|})/(2\sqrt{|N|} - 1)$. In the first step this is obviously true,
    as any MMS partition of the selected agent can be used. In any other step,
    we will use one of two possible strategies to divide the goods. Note that
    since each step allocates one or more bundles from a feasible allocation,
    there at any point remains at most as many goods from each category as can
    be allocated to the remaining agents.

    First, notice that when $b$ bundles have been allocated, the remaining value
    is at least $|N|\mu_i - b\alpha\mu_i$ for any remaining agent $i$.
    By \cref{lem:correctness}, \cref{alg:approx} can be used to find a partition
    of the remaining goods so that the value of each bundle is at least a $(|N|
    - b)/(2(|N| - b) - 1)$ share of the remaining value, given that any single
    good is not worth more than this share. If the remaining value is at least
    $(|N| - b - 1)2\alpha\mu_i + \alpha\mu_i$, then this guarantees that the
    method gives each remaining agent a bundle of value at least $\alpha\mu_i$.
    Through valid reductions, the condition on the maximum value of any
    individual good can easily be achieve before the lone-divider technique is
    applied. Note that when \cref{alg:approx} is applicable, a final allocation
    can immediately be achieved, rather than having to following the
    lone-divider strategy any further. Thus, we only wish to show that the
    lone-divider strategy can be performed until the remaining value to agent
    ratio is high enough.

    If $\alpha = (\sqrt{|N|})/(2\sqrt{|N|} - 1)$, the remaining value may
    initially be less than $2(|N| - b - 1)\alpha\mu_i + \alpha\mu_i$. However, as
    long as $b\alpha\mu_i \le (b + 1)(1 - \alpha)\mu_i$, the partition created
    by taking one of $i$'s MMS partitions and removing all already allocated
    goods, contains at least $|N| - b$ bundles with a value of at least
    $\alpha\mu_i$. This partition can contain more than $|N| - b$ non-empty
    bundles. To turn it into a partition that contains exactly $|N| - b$ bundles
    of value at least $\alpha\mu_i$, select a subset of $|N| - b$ bundles, each
    with a value of at least $\alpha\mu_i$. For any other bundle, transfer the
    goods to any one of the bundles in the selected subset that has space for
    more goods. There is always a bundle with space left, as there are at most
    as many remaining goods in each category as can fit in $|N| - b$ bundles.

    If for some remaining agent $i$, $b\alpha\mu_i \le (b + 1)(1 -
    \alpha)\mu_i$, then another step of the lone-divider strategy can be
    performed by selecting $i$. We thus need to show that for $\alpha =
    (\sqrt{|N|})/(2\sqrt{|N|} - 1)$, when $b\alpha\mu_i > (b + 1)(1 -
    \alpha)\mu_i$ for each remaining agent $i$, then $|N|\mu_i - b\alpha\mu_i
    \ge (|N| - b - 1)2\alpha\mu_i + \alpha\mu_i$ for all remaining agents $i$.
    If $\mu_i = 0$, then $b\alpha\mu_i = 0 = (b + 1)(1 - \alpha)\mu_i$. So, we
    can without loss of generality assume that $\mu_i = 1$ for all remaining
    agents $i$ when $b\alpha\mu_i > (b + 1)(1 - \alpha)\mu_i$. Since $\alpha >
    1/2$, we have the following when $b\alpha > (b + 1)(1 - \alpha)$:
    \begin{align*}
        b\alpha &> (b + 1)(1 - \alpha) \\
        2b\alpha - b &> 1 - \alpha \\
        b &> \frac{1 - \alpha}{2\alpha - 1}
    \end{align*}

    \noindent We now need to show that when $b > \frac{1 - \alpha}{2\alpha -
    1}$, then for $\alpha = (\sqrt{|N|})/(2\sqrt{|N|} - 1)$ we have $|N| -
    b\alpha \ge (|N| - b - 1)2\alpha + \alpha$. That is, we need to show that
    \begin{align*}
        |N| - b\alpha &\ge (|N| - b - 1)2\alpha + \alpha \\
        |N| &\ge (2|N| - b - 1)\alpha \\
            &\ge \left(2|N| - \frac{1 - \alpha}{2\alpha - 1} - 1\right)\alpha \\
            &\ge 2|N|\alpha - \frac{\alpha - \alpha^2}{2\alpha - 1} -
            \frac{2\alpha^2 - \alpha}{2\alpha - 1} \\
            &\ge 2|N|\alpha - \frac{\alpha^2}{2\alpha - 1}
    \end{align*}

    \noindent Which can be reorganized to $4\alpha|N| - |N| \ge (4|N| -
    1)\alpha^2$. Using $\alpha = \smash{\frac{\sqrt{|N|}}{2\sqrt{|N|} - 1}}$, we
    get that
    \begin{align*}
        4\alpha|N| - |N| &= \frac{4|N|\sqrt{|N|}}{2\sqrt{|N|} - 1} - |N| \\
            &= \frac{(4|N|\sqrt{|N|})(2\sqrt{|N|} - 1)}{(2\sqrt{|N|} -
            1)^2} - \frac{|N|(2\sqrt{|N|} - 1)^2}{(2\sqrt{|N|} - 1)^2} \\
            &= \frac{8|N|^2 - 4|N|\sqrt{|N|}}{(2\sqrt{|N|} - 1)^2} -
            \frac{4|N|^2 - 4|N|\sqrt{|N|} + |N|}{(2\sqrt{|N|} - 1)^2} \\
            &= \frac{4|N|^2 - |N|}{(2\sqrt{|N|} - 1)^2} \\
            &= \frac{(4|N| - 1)|N|}{(2\sqrt{|N|} - 1)^2} \\
            &= (4|N| - 1)\left(\frac{\sqrt{|N|}}{2\sqrt{|N|} - 1}\right)^2 \\
            &= (4|N| - 1)\alpha^2
    \end{align*}

    \noindent Since the left and right hand sides are equal, $4\alpha|N| - |N|
    \ge (4|N| - 1)\alpha^2$ and setting $\alpha = (\sqrt{|N|}))/(2\sqrt{|N|} - 1)$
    guarantees that at least one of the two methods work. Consequently, a
    ($\sqrt{|N|}/(2\sqrt{|N|} - 1)$)-approximate MMS allocation always exists.
    Since the two sides are equal for all possible values of $|N|$, this is the
    greatest $\alpha$ this method works for.
    \qed
\end{proof}

\section{Omitted Proofs from \cref{sec:uniform-matroid-constraints}}

\begin{proof}[for \cref{lem:single-category-value-adjustment}]
    Let $A$ be an MMS partition of $i$ and $S = \bigcup_{A_k \in A^*}A_k$, where
    $A^*$ is a set of $|N| - r + 1$ bundles in $A$ with $A_k \cap \{1, 2, \dots,
    r - 1\} = \emptyset$ for each $A_k \in A^*$. Since $0 < r \le |N|$ and at
    most $r - 1$ bundles can contain goods from $\{1, 2, \dots, r - 1\}$, there
    are at least $|N| - (r - 1) = |N| - r + 1$ bundles that can be used in $A^*$.
    Further, there is no good $j \in (M \setminus (\{1, 2, \dots, r - 1\} \cup
    B_r))$ with $v_i(j) > v_i(j')$ for any good $j' \in B_r$. Thus, we have that
    for $j \in S$, either $j \in B_r$ or $v_i(j') \ge v_i(j)$ for all $j' \in B
    \setminus S$.  As either $|B_r| = k_1(|N| - r + 1)$ or $|B_r| = |M \setminus
    \{1, 2, \dots, r - 1\}|$, we have $|B_r| \ge |S|$, and
    \[
        v_i(B_r) \ge v_i(S) = \sum_{A_k \in A^*} v_i(A_k) \ge \sum_{A_k \in
        A^*} \mu_i = (|N| - r + 1)\mu_i
    \]
    \qed
\end{proof}

\begin{proof}[for \cref{lem:alg-2/3-MMS-correctness}]
    To prove that the algorithm finds a $(2/3)$-MMS allocation in polynomial
    time, we wish to show that
    \begin{inl}
        \item\label{item:2/3-MMS-condition-1}any $B'_j$ allocated is feasible,
        \item\label{item:2/3-MMS-condition-2}all goods are allocated,
        \item\label{item:2/3-MMS-condition-3}both while-loops
            (lines~\ref{line:2/3-MMS-item-adding} and
            \ref{line:2/3-MMS-item-upgrade}) finish running for each $B'_j$, and
        \item\label{item:2/3-MMS-condition-4}the algorithm finishes in
            polynomial time.
    \end{inl}
    Note that \ref{item:2/3-MMS-condition-3} guarantees that there is always an
    agent that values $B'_j$ at $2/3$. As a consequence, as long as
    \ref{item:2/3-MMS-condition-3} holds, each agent $i \in N$ will receive a
    bundle worth $(2/3)\mu_i$.
    \\
    \vspace{0.5\baselineskip}

    \noindent \textbf{Feasible and complete allocation}
    \\
    \vspace{0.2\baselineskip}

    \noindent For \ref{item:2/3-MMS-condition-1} and
    \ref{item:2/3-MMS-condition-2}, we wish to show that at the end of each
    iteration of the for loop, $|M| \le k_1|N|$. By the assumption that any
    instance has at least one feasible allocation, this holds before the first
    iteration. Assume that this holds at the start of a iteration for a specific
    $j$, i.e., $|M| \le jk_1$. Then, we must show that $|M| - |B'_j| \le (j -
    1)k_1$ when $B'_j$ is allocated to an agent. Throughout an iteration, the
    size of $B'_j$ never decreases. If line~\ref{line:2/3-MMS-minimum-goods} is
    executed, then $|B'_j| > 1 + (|M| - k_1(j - 1) - 1) = |M| - k_1(j - 1)$ and
    it follows that $|M| - |B'_j| < |M| - (|M| - k_1(j - 1)) = k_1(j - 1)$. If
    on the other hand, line~\ref{line:2/3-MMS-minimum-goods} is not executed,
    then $|M| \le k_1(j - 1) + 1 = k_1(j - 1) + |B'_j|$. Consequently, $|M| -
    |B'_j| \le (j - 1)k_1$ always holds. After the last iteration, $|N| = 0$
    implying that $|M| \le 0k_1 = 0$. Thus, \ref{item:2/3-MMS-condition-2} must
    hold as long as \cref{alg:2/3-MMS} finishes. Since the loop on
    line~\ref{line:2/3-MMS-item-adding} only adds items as long as $|B'_j| <
    k_1$ and the loop on line~\ref{line:2/3-MMS-item-upgrade} does not change
    $|B'_j|$, $|B'_j| \le k_1$ as long as line~\ref{line:2/3-MMS-minimum-goods}
    does not add too many goods to $B'_j$. However, since $|M| \le jk_1$, the
    line adds at most $jk_1 - k_1(j - 1) - 1 = k_1 - 1$ goods to $B'_j$, which
    at this point only contains a single good. Thus,
    \ref{item:2/3-MMS-condition-1} holds.
    \\
    \vspace{0.5\baselineskip}

    \noindent \textbf{Bundle value always reaches $\mathbf{2/3}$}
    \nopagebreak
    \\
    \vspace{0.2\baselineskip}

    \noindent For \ref{item:2/3-MMS-condition-3} we wish to show that during the
    execution of the first loop (line~\ref{line:2/3-MMS-item-adding}), either
    $v_i(B'_j) \ge 2/3$ for some $i \in N$ or $|B'_j| = k_1$. Additionally, we
    wish to show that during the second loop, $v_i(B'_j) \ge 2/3$ for some $i
    \in N$. In order to show this, let $\smash{B_r^j}$ denote the bundle
    consisting of the $\min(|B_r \cap M|, k_1(j - r + 1))$ most valuable
    remaining goods in $B_r$ (i.e., in $ B_r \cap M$) at the start of
    iteration $j$ for all $r \le j$. We wish to show that in any iteration
    $j$, for all $i \in N$ we have $v_i(\smash{B_r^j}) \ge j - r + 1$.
    This will in turn guarantee that $v_i(\smash{B_r^j}) \ge 1$ for all
    remaining agents $i$ when $j = r$, and the bundle $B'_j$ will be able to
    reach a value of $2/3$ either in the first loop when $|M \setminus (B'_1
    \cup B'_2 \cup \dots \cup B'_{j - 1})| \le k_1$, or otherwise by the time
    the second loop runs out of improvements to make, as $|B_r^r| \le
    k_1$.

    By \cref{lem:single-category-value-adjustment}, it follows that when $j =
    |N|$ in the first iteration, $B_r = \smash{B_r^{j}}$. Thus,
    $v_i(\smash{B_r^{j}}) = v_i(B_r) \ge j - r + 1$ and the invariant holds
    initially. Now assume that $v_i(\smash{B_r^j}) \ge j - r + 1$ for some $j >
    1$ and all $r \le j$. We wish to show that the invariant then also holds for
    $j' = j - 1$. In other words, we wish to show that $v_i(\smash{B_r^{j'}}) >
    (j' - r + 1)$ for all $r \le j'$ and remaining agents $i$. There are two
    possible situations that need to be accounted for, depending on if
    $v_i(B'_j) > 1$ or $v_i(B'_j) \le 1$ for agent $i$.

    Since any good $g \in (M \setminus (B'_1 \cup B'_2 \cup \dots \cup B'_j))$
    has $v_i(g) < 1/3$, the only way for $v_i(B'_j)$ to be greater than $1$ is
    if this occurs when the bundle is modified on
    line~\ref{line:2/3-MMS-minimum-goods}. In that case, the bundle will not be
    modified further before it is allocated and we have that $|M \setminus B'_j|
    = j'k_1 \implies |\smash{B_r^{j'}}| = k_1(j' - r + 1)$. Further, since
    $B'_j$ consist of $j$ and the at most $k_1 - 1$ least valuable goods in $M$,
    we know that $\smash{B_r^{j'}}$ consists of, in addition to $\{r, r + 1,
    \dots j'\}$, $(k_1 - 1)(j' - r + 1)$ goods in $M$, each with at least the
    same value as any of the maximum $k_1 - 1$ goods in $B'_j \setminus \{j\}$.
    In other words,
    \begin{align*}
        v_i(B_r^{j'})
            &= v_i(\{r, r + 1, \dots, j'\}) + v_i(B_r^{j'} \setminus
            \{r, r + 1, \dots, j'\}) \\
            &\ge (j' - r + 1)v_i(j) + (j' - r + 1)v_i(B'_j \setminus
            \{j\}) \\
            &= (j' - r + 1)v_i(B'_j) \\
            &> j' - r + 1
    \end{align*}

    \noindent This leaves the case where $v_i(B'_j) \le 1$. In this case we distinguish
    between two cases, depending on if $\smash{B_r^{j'}} = \smash{B_r^{j}}
    \setminus B'_j$ or not. If this holds true, then
    \[
        v_i(B_r^{j'}) = v_i(B_r^{j} \setminus B'_j) \ge v_i(B_r^{j}) -
        v_i(B'_j) > (j - r + 1) - 1 = j' - r + 1
    \]

    \noindent If $\smash{B_r^{j'}} \neq \smash{B_r^{j}} \setminus B'_j$,
    we claim that $B'_j \setminus \{j\}$ does not contain any good better than
    the $(j - r)k_1 + 3$ most valuable good in $\smash{B_r^j}$. If a good
    $g$, that is the $(j - r)k_1 + 2$ most valuable good in
    $\smash{B_r^j}$ or better, is added to $B'_j$ on
    line~\ref{line:2/3-MMS-minimum-goods} or in the first loop
    (line~\ref{line:2/3-MMS-item-adding}), then $B'_j$ contains all goods in $M$
    that are worse than $g$, and by definition all goods in $\smash{B_r^{j}}$
    that are worse than $g$. This is a contradiction, as then
    $|\smash{B_r^{j}} \setminus B'_j| \le k_1(j - 1)$ and $M \setminus (B'_1
    \cup B'_2 \cup \dots \cup B'_j) = \smash{B_r^{j}} \setminus B'_j$, which
    implies that $\smash{B_r^{j'}} = \smash{B_r^{j}} \setminus B'_j$.
    Thus, the only way that such a $g$ could be added to $B'_j$ and maintain
    $\smash{B_r^{j'}} \neq \smash{B_r^{j}} \setminus B'_j$, is in the second
    loop (line~\ref{line:2/3-MMS-item-upgrade}). However, when a good $g$ is
    added to $B'_j$ in the second loop, $|B'_j| = k_1$ and $B'_j$ contains the
    goods $\{g + 1, g + 2, \dots, g + k_1 - 2\}$. Thus, since $g$ is the $(j -
    r)k_1 + 2$ most valuable good in $\smash{B_r^{j}}$ or better, $g + k_1 - 2
    \in \smash{B_r^{j}}$ as $\smash{B_r^{j}}$ by definition contains as many
    goods as possible up to $(j - r + 1)k_1$ goods and $g + k_1 - 2$ is the $(j
    - r)k_1 +
    2 + k_1 - 2 = (j - r + 1)k_1$ good in $\smash{B_r^{j}}$.  Consequently,
      $B'_j \in \smash{B_r^{j}}$ and $\smash{B_r^{j'}} = \smash{B_r^{j}}
      \setminus B'_j$ which is another contradiction and the claim holds.

    It remains to show that when $B'_j$ contains $j$ and no better than the $(j
    - r)k_1 + 3$ good in $\smash{B_r^j}$, then $v_i(\smash{B_r^{j'}})
    \ge (j' - r + 1)$. We can divide $\smash{B_r^j}$ into $j - r + 1$
    bundles of at most $k_1$ goods each, by creating the bundles $\{r\},
    \{r + 1\}, \dots, \{j\}$ and then as long as there remains goods in
    $\smash{B_r^j}$, placing the most valuable remaining good into the first
    of the bundles which does not have $k_1$ goods yet. Then the bundle that
    started with the good $j$ contains, except for $j$, no good better than the
    $(j - r)k_1 + 2$ most valuable good in $\smash{B_r^j}$ and since $j$
    is the least valuable good of the ones initially placed in the bundles, the
    last bundle is the least valuable of any of the bundles. In other words, the
    value of all but the last bundle is at least $((j - r)/(j - r +
    1))v_i(\smash{B_r^j}) = j - r = j' - r + 1$.  Since $B'_j$ only
    intersects this last bundle, and there are no more than $(j - r)k_1$ goods
    in the other bundles, there remains at least $(j' - r + 1)k_1$ goods in
    $\smash{B_r^j}$ with a combined value of at least $j' - r + 1$ and the
    invariant holds for $j'$. By induction it holds for all values of $j$ and
    \ref{item:2/3-MMS-condition-3} holds.
    \\
    \vspace{0.5\baselineskip}

    \noindent \textbf{Polynomial run time}
    \\
    \vspace{0.2\baselineskip}

    \noindent It remains to show that \ref{item:2/3-MMS-condition-4} holds,
    namely that the algorithm uses polynomial time. The outer loop has as many
    iteration as there are agents, so it suffice to show that each iteration is
    polynomial. The loop on line~\ref{line:2/3-MMS-item-adding} runs for at most
    $k_1 - 1$ iterations, as $|B'_j| \ge 1$ prior to the first iteration and each
    iteration increases the size of $B'_j$ by $1$ up to a maximum of $k_1$. The
    loop on line~\ref{line:2/3-MMS-item-upgrade} also runs for a maximum of
    $|M|$ iterations, as each iteration exchanges the least valuable good $g \in
    B'_j$ for the least valuable, but better good in $M \setminus (B'_1 \cup
    B'_2 \cup \dots \cup B'_j)$. Since $g$ now is worse than all goods in
    $B'_j$, it will never be picked again. Thus, the size of the set of goods in
    $M \setminus (B'_1 \cup B'_2 \cup \dots \cup B'_j)$ that will never be
    picked again increases by $1$ each iteration and after (less than) $|M|$
    iterations there is no good to pick. Since each loop has a polynomial number
    of iterations and each individual operation can be performed in polynomial
    time, the algorithm must finish in polynomial time and
    \ref{item:2/3-MMS-condition-4} holds.
    \qed
\end{proof}

\begin{proof}[of \cref{thr:mms-2/3-single-category}]
    We wish to show that for any instance $I$ we can in polynomial time either
    convert the instance into one that \cref{alg:2/3-MMS} accepts or failing
    that we can directly create a $(2/3)$-approximate MMS allocation.

    We can by \cref{thr:ordered} convert $I$ to an ordered
    instance. The other requirements for \cref{alg:2/3-MMS} can be achieved by
    performing the following steps. Note that
    step~\ref{step:2/3-single-category-3} may cause some remaining agent $i'$ to
    have $v_{i'}(M) = 0$, in which case step~\ref{step:2/3-single-category-1}
    cannot possibly rescale the agent's valuations so that $v_{i'} = |N|$.  In
    this case, just skip the rescaling of agent $i'$'s valuations, and agent
    $i'$ will at some point be reduced away in
    step~\ref{step:2/3-single-category-3}. The same may happen for $i'$ and
    step~\ref{step:2/3-single-category-2} if $v_{i'}(|N|) = 0$. In which case
    step~\ref{step:2/3-single-category-3} will find a valid reduction with $i'$.
    \begin{enumerate}
        \item\label{step:2/3-single-category-1} For all $i \in N$, rescale $i$'s
            valuations so that $v_i(M) = |N|$.
        \item\label{step:2/3-single-category-2} If for any $i \in N, r \in
            \{1, 2, \dots, |N|\}$, $v_i(B_r) < |N| - r + 1$, rescale $i$'s
            valuations so that $v_i(B_r) = |N| - r + 1$.
        \item\label{step:2/3-single-category-3} If $v_i(1) \ge 2/3$ or
            $v_i(\{|N|, |N| + 1\}) \ge 2/3$ for $i \in N$, construct a valid
            reduction with, respectively, $\{1\}$ or $\{|N|, |N| + 1\}$ and
            agent $i$, and go back to step~\ref{step:2/3-single-category-1}.
    \end{enumerate}

    \noindent First, we wish to show that after
    step~\ref{step:2/3-single-category-2}, it holds that $\mu_i \le 1$ and
    $v_i(B_r) \ge |N| - r + 1$ for all $i$ and $r$. By
    \cref{thr:normalized} it holds that $\mu_i \le 1$ for all $i \in N$ after
    step~\ref{step:2/3-single-category-1}. In
    step~\ref{step:2/3-single-category-2}, the rescaling increases the value of
    all goods (of non-zero value). Thus, rescaling for a specific $i$ and
    $r$ does not decrease the value of $v_i(B_{r'})$ for $r' \in \{1,
    2, \dots, |N|\}$. Hence, after step~\ref{step:2/3-single-category-2} it
    holds that $v_i(B_r) \ge |N| - r + 1$ for all $i \in N$, $r \in \{1, 2,
    \dots, |N|\}$. By \cref{lem:single-category-value-adjustment}, we know that
    $v_i(B_r) \ge (|N| - r + 1)\mu_i$. Since each rescaling sets $v_i(B_r) = |N|
    - r + 1$, it follows that $\mu_i \le 1$.

    Since $\mu_i \le 1$,
    \cref{cor:single-item-reduction,cor:double-item-reduction} guarantee that if
    one of the conditions in step~\ref{step:2/3-single-category-3} hold, a valid
    reduction can be created for $\alpha = 2/3$. Thus, it also holds that when
    no valid reduction is found in step~\ref{step:2/3-single-category-3}, then
    for all $i \in N$ we have $v_i(1) < 2/3$ and $v_i(\{|N|, |N| + 1\}) < 2/3
    \implies v_i(|N| + 1) < 1/3$.

    The only missing condition of \cref{alg:2/3-MMS} is $|M| > |N|$. If $|M| \le
    2|N|$ at any point, it must hold for any $i \in N$ that either $v_i(1) \ge
    2/3$ or $v_i(|N| + 1) \ge 1/3$ when $v_i(M) \ge |N|$ and $|M| > 0$.
    Therefore, when step~\ref{step:2/3-single-category-3} finishes without going
    back to step~\ref{step:2/3-single-category-1}, either $|M| = 0$ or
    $|M| > 2|N| > |N|$. In the first case, we already have a $(2/3)$-approximate
    MMS allocation and in the latter case the missing condition of
    \cref{alg:2/3-MMS} holds for the instance.

    Since both $r$ and $i$ are bounded in the number of agents, it can easily
    be verified that each individual step can be performed in polynomial time.
    As each valid reduction removes an agent, the number of times the steps are
    performed is also bound in the number of agents. The preprocessing can
    therefore be done in polynomial time.
    \qed
\end{proof}

\begin{proof}[for \cref{lem:few-items-reduction-to-unconstrained}]
    Any feasible allocation for an instance $I$ of the fair allocation problem
    under cardinality constraints is also a feasible allocation for the
    unconstrained instance $I' = \langle N, M, V \rangle$. By definition we thus
    have that $\mu_i^{I'} \ge \mu_i^{I}$. If no agent $i \in N$ has $\mu_i^I =
    0$, then for any $\alpha > 0$, any $\alpha$-approximate MMS allocation of
    $I'$ must allocate at least one good to each agent $i$. Consequently, no
    agent may receive more than $|M| - |N| + 1 < |N| + k_1 - |N| + 1 = k_1 + 1$
    goods.  Any $\alpha$-approximate MMS allocation of $I'$ is then a feasible
    $\alpha$-approximate MMS allocation for $I$.

    By \cref{thr:ordered} we can assume that $I$ is ordered. If $\mu_i = 0$ for
    an agent $i \in N$, then $v_i(j) \ge \mu_i$ for all $j \in M$ and
    \cref{cor:single-item-reduction} can be used to reduce away $i$. Since the
    reductions remove one item along with every agent removed, the conditions of
    the lemma still hold for the reduced instance. The check can be performed in
    polynomial time, as $\mu_i = 0 \Leftrightarrow v_i(|N|) = 0$.
    \qed
\end{proof}

\begin{proof}[for \cref{lem:single-category-low-threshold}]
    To show the existence of an MMS allocation when $k_1 \le 2$, we wish to show
    that the instance $I$ can be reduced to an ordered instance $I' = \langle
    N', M', V', \langle (C_1', k_1) \rangle \rangle$ with $|M'| = 2|N'|$, in which
    the allocation $A = \langle \{1, 2|N'|\}, \{2, 2|N'| - 1\}, \dots, \{|N'|,
    |N'| + 1\}$ is an MMS partition for all agents in $I'$.

    By \cref{thr:ordered} we can assume that $I$ is ordered. If $|M| < 2|N|$,
    then in any allocation there is at least one bundle containing only a single
    item. In other words, \cref{cor:single-item-reduction} allows for a valid
    reduction with any agent $i \in N$ and the good $1$ for $\alpha = 1$. Thus,
    repeated reductions can be performed until we have an instance $I'$ where
    either $|M'| = 0$ and we have found an MMS allocation or $|N'| > 0$ and
    $|M'| = 2|N'|$. In the second case, for any agent $i \in N'$ let $A'$ be any
    MMS partition of $I$ for $i$. We wish to show that $A'$ can be turned into
    $A$ without reducing the value of the least-valuable bundle in $A'$.

    Let $g$ be the first good in $\{1, 2, \dots, |N|\}$ such that $\{g, 2|N'| -
    g + 1\} \notin A'$. Then $A'$ contains distinct bundles $B_g = \{g, g'\}$
    and $B_{2|N'| - g + 1} = \{g'', 2|N'| - g + 1\}$. Since $g$ was selected to
    be the smallest $g$ for which this holds, all less-valuable goods than
    $2|N'| - g + 1$ and more valuable than $g$ appear in other bundles than
    $B_g$ and $B_{2|N'| - g + 1}$. Consequently, we have $v_i(g') \ge v_i(2|N'|
    - g + 1)$, $v_i(g'') \ge v_i(2|N'| - g + 1)$ and $v_i(g) \ge v_i(g'')$.
    Thus, $v_i(\{g'', g'\}) \ge v_i(B_{2|N'| - g + 1})$ and $v_i(\{g, 2|N'| - g
    + 1\}) \ge v_i(B_{2|N'| - g + 1})$. We can swap the location
    of $g'$ and $2|N'| - g + 1$ to create an allocation where the worst bundle
    is no worse than $B_{2|N'| - g + 1}$ and that shares one more bundle with
    $A$. This can be repeated until the allocation shares all its bundles with
    $A$. In other words, $A$ is an MMS partition of $i$. Since all agents share
    the same MMS partition, it is also an MMS allocation.

    For $k_1 \ge 4$, we will show that there for any $k_1 \ge 4$ exists an
    instance of the problem for which no MMS allocation exists. The instance
    will be created by introducing cardinality constraints to the unconstrained
    instance of \citeauthor{feige_tight_2021} with $3$ agents and $9$ goods for
    which they showed that the best possible allocation achieves no more than an
    approximation ratio of $39/40$ \cite{feige_tight_2021}.
    \citeauthor{feige_tight_2021}'s proof used MMS partitions that contained no
    more than four goods in each bundle. By introducing cardinality constraints
    with a single category and a threshold of $k_1 \ge 4$, these MMS partitions
    remain feasible and each agent's MMS stays the same. By introducing
    cardinality constraints, the set of feasible allocations is a subset of the
    set of allocations for the unconstrained instance. Consequently, no feasible
    allocation can provide all the agent's with more value than the best
    allocation in the unconstrained instance, and there does not exist any MMS
    allocation.
    \qed
\end{proof}

\section{Omitted Proof from \cref{sec:chores}}

\begin{proof}[for \cref{thr:scale-invariance-chores}]
    Exactly as the proof of \cref{thm:scale-inv}.
    \qed
\end{proof}

\begin{proof}[for \cref{thr:normalization-chores}]
    Exactly as the proof of \cref{thr:normalized}, except that the $1$ in the last
    equation is exchanged for $-1$.
    \qed
\end{proof}

\begin{proof}[for \cref{thr:ordered-instance-reduction-chores}]
    Exactly as the proof of \cref{thr:ordered}.
    \qed
\end{proof}

\begin{proof}[for \cref{thr:value-adjustment-chores}]
    Let $B = \{g_{i_{|N|r + 1 - r}}, g_{i_{|N|r + 2 - r}}, \dots, g_{i_{|N|r +
    1}}\}$. By the pigeonhole principle, at least one bundle $A_j$ in $i$'s MMS
    partition must contain at least $r + 1$ chores from $\{g_{i_1}, g_{i_2},
    \dots, g_{i_{|N|r + 1}}\}$. Since $B$ contains the $r + 1$ least valuable
    goods in this set, we have $v_i(B) \ge v_i(A_j)
    \ge \mu_i$.
    \qed
\end{proof}

\subsection{Proof for \cref{thr:2-MMS-chores}}

To prove \cref{thr:2-MMS-chores}, we will show that \cref{alg:2-MMS-chores}, a
variation of \cref{alg:approx}, finds $(|N|/(2|N| - 1))$-approximate MMS
allocations for ordered instances where no chore is worth less than $-1$, $\mu_i
\le -1$ and $v_i(M) > -|N|$ for all $i \in N$. \Cref{alg:2-MMS-chores} works in
a similar manner to \cref{alg:approx}, it starts by creating a bundle $B$
consisting of the $\lceil |C_h|/n \rceil$ worst chores in each category $C_h$
(denoted by $C_h^H$). It then gradually, as to not improve the value of the
bundle by more than $1$, improves the value of the bundle by exchanging a good
in some $C_h^H$ for one of the $\lceil |C_h|/n \rceil$ best chores in the same
category (denoted by $C_h^L$). To mitigate the effects of rounding $C_h/n$, it
can also remove the $\lceil |C_h|/n \rceil$ best chore in any $C_h$ where $\lceil
|C_h|/n \rceil > |C_h|/n$. This strategy guarantees, as for goods, that the bundle
created is feasible and that there at any point remains at most as many chores
as can be allocated to the remaining agents. Additionally, it makes sure that
the bundle initially contains at least $1/n$ of the remaining disutility and in
the end at most $1/n$ of the remaining disutility. Thus, a similar argument can
be made about the upper bound on the remaining disutility, as for the lower
bound on remaining value for goods.

\begin{figure}[t]
\begin{procedure}{Find a $\alpha$-MMS solution to ordered chore instance}{2-MMS-chores}
\Preamble{An ordered instance $I = \langle N, M, V, C \rangle$ with all $v_{ij}
    \ge -1$, $v_i(M) > -|N|$ and $\mu_i \le -1$}{Allocation $A$ consisting of
    each bundle $B$ allocated}
\begin{pseudo}
    \kw{while} there is more than one agent left \\+
        $B$ = $\cup_{h = 1}^{\ell} C_h^H$ \\
        \kw{while} $v_i(B) < -\alpha$ for all agents $i$ \\+
            \kw{if} $B \cap C_h^H \neq \emptyset$ for some $C_h$ \\+
                $j =$ any element of $C_h^L \setminus B$ \\
                $j' =$ any element of $B \cap C_h^H$ \\
                $B = (B \setminus \{j'\})\cup \{j\}$ \\-
            \kw{else} $j =$ any $c_{\lceil |C_h|/n \rceil}$ in $B$ for $C_h$
            with $|C_h|/n < \lceil |C_h|/n \rceil$ \\+
                $B = B \setminus \{j\}$ \\--
        allocate $B$ to some agent $i$ with $v_i(B) \ge -\alpha$ \\
        remove $B$ and $i$ from $I$ and update $n$, and $C_h^H$ and $C_h^L$ for
        all $h$ \\-
    allocate the remaining chores to the last agent
\end{pseudo}
\end{procedure}
\end{figure}

\begin{lemma}\label{lem:remaining-value-chores}
    Let $I = \langle N, M, V, C \rangle$ be an ordered instance of the
    fair allocation of chores problem under cardinality constraints where each
    chore is worth no less than $-1$ and $v_i(M) \ge -|N|$ for each $i \in N$.
    Let $n$ denote the number of remaining agents at any point during the
    execution of \cref{alg:2-MMS-chores}. Then for any $\alpha \in (1, 2]$, each
    remaining agent assigns a value of at least $-|N| + (|N| - n)(\alpha - 1)$
    to the set of unallocated chores at any point during the execution of the
    algorithm.
\end{lemma}

\begin{proof}
    Since $v_i(M) \ge -|N|$, this holds at the start of the algorithm. Assume
    that there are $n$ remaining agents at the start of an iteration and
    for each remaining agent $i$, $v_i(M) \ge -|N| + (|N| - n)(\alpha - 1)$. Let
    $i'$ be the agent receiving $B$ in the iteration. For any remaining agent $i
    \neq i'$, we wish to show that $v_i(M \setminus B) \ge -|N| + (|N| - n +
    1)(\alpha - 1)$. Due to the additive valuations, the only way that $v_i(M
      \setminus B) < -|N| + (|N| - n + 1)(\alpha - 1)$ is if $v_i(B) > -\alpha +
      1$. Since any change to $B$ after the initial creation removes a chore
      from $B$ or exchanges a chore in $B$ for another, any individual change
      cannot increase the value of $B$ by more than $1$. Thus, the only way for
      $v_i(B) > -\alpha + 1$ is if $B = \smash{\bigcup_{h = 1}^\ell} C_h^H$ and
      $v_i(\smash{\bigcup_{h = 1}^\ell} C_h^H) > -\alpha + 1$. However, by
      definition $v_i(C_h^H) \le v_i(C_h)/n$ which implies $v_i(B) \le
      v_i(M)/n$.  Consequently, $v_i(M \setminus B) \ge (n - 1)v_i(B) > -(n -
      1)(\alpha - 1) \ge -|N| + |N|(\alpha - 1) - (n - 1)(\alpha - 1) = -|N| +
      (|N| - n + 1)(\alpha - 1)$.
    \qed
\end{proof}

\noindent With \cref{lem:remaining-value-chores} we have a sufficient upper
guarantee for the remaining disutility. We are now ready to show the guarantees
of \cref{alg:2-MMS-chores}.

\begin{lemma}\label{lem:correctness-chores}
    Given a normalized ordered instance $I = \langle N, M, V, C \rangle$ of the
    fair allocation problem under cardinality constraints where $\mu_i \le -1$,
    $v_{ij} \ge -1$ and $v_i(M) > -|N|$ for all $i \in N$, $j \in M$, and
    $\alpha = (2|N| - 1)/|N|$, \cref{alg:2-MMS-chores} finds a feasible $(2|N| -
    1)/|N|$-approximate MMS allocation in polynomial time in the number of
    agents and chores.
\end{lemma}

\begin{proof}
    When allocating the remaining chores to the last agent,
    \cref{lem:remaining-value-chores} guarantees that the chores are worth at
    least $-\alpha$, if $-|N| + (|N| - 1)(\alpha - 1) \ge -\alpha$, which holds for
    $\alpha \ge (2|N| - 1)/|N|$. Additionally, as long as $B$ reaches a value of
    $-\alpha$ before running out of improvement operations, any other agent is
    also guaranteed to receive a bundle they value at no less than $-\alpha$.
    Since $B$ contains the $\lfloor C_h/n \rfloor$ best chores in each category
    $C_h$ when the algorithm runs out of operations, $B$ will contain chores of
    no more than $1/n$ of the remaining disutility. We thus only need to show
    that the remaining value is always at least $-n\alpha$ for any remaining
    agent. \Cref{lem:remaining-value-chores} guarantees that the remaining value
    is at least $-|N| + (|N| - n)\alpha$. Since, this is at least $-\alpha$ for
    $n = |N| - 1$ for $\alpha \ge (2|N| - 1)/|N|$, the value is at least $-(n -
    1)(\alpha - 1) - \alpha \ge -n\alpha$ for any other $n$, and we are
    guaranteed that the value of $B$ reaches at least $-\alpha$ in any
    iteration. Since $\mu_i \le -1$ for $i \in N$, each agent $i$ receives at
    least $-\alpha\mu_i$ value.

    It remains to show that any bundle allocated is feasible. As long as $|C_h|
    \le nk_h$, it holds that $\lceil |C_h|/n \rceil \le k_h$ and any bundle
    allocated is feasible. Obviously, $|C_h| \le nk_h$ holds when $n = |N|$, as
    all instances are assumed to have at least one feasible complete allocation.
    Assume that $|C_h| \le nk_h$ holds at the start of an iteration. The bundle
    $B$ contains at least $\lfloor |C_h|/n \rfloor \ge |C_h| - (n - 1)k_h$ of
    the chores in $C_h$ at any point during an iteration. Thus, $|C_h \setminus
    B| \le (n - 1)k_h$ and the condition holds for $n - 1$ after allocating $B$.
    Consequently, each allocated bundle, including the bundle allocated to the
    last agent, is feasible. Since the last agent receives all remaining chores,
    all chores are allocated.

    In each iteration of the algorithm, chores are removed from $B$ and
    exchanged through a set of operations. As each chore is not added back into
    $B$ after being removed, the number of operations in each iteration is
    polynomial in the number of agents and chores. Since there are $|N| - 1$
    iterations, the running time of the algorithm is also polynomial in the
    number of agents and chores.
    \qed
\end{proof}

\noindent We can now combine \cref{lem:correctness} with rescaling of valuations
in order to show that $((2|N| - 1)/|N|)$-approximate MMS allocations always
exist and can be found in polynomial time.

\begin{proof}[for \cref{thr:2-MMS-chores}]
    First of all, if the instance $I$ has any agent $i \in I$ with $v_i(M) = 0$,
    then we know we can remove $i$ from the instance by allocating the
    $k_h$ worst chores in each $C_h$ to $i$. Thus, we can assume $v_i(M) < 0$.

    The instance can by \cref{thr:ordered-instance-reduction-chores} easily be
    turned into an ordered instance. Further, the valuations of each agent $i$
    can be rescaled so that $v_i(M) = -|N|$, which by
    \cref{thr:normalization-chores} gurantees that $\mu_i \le -1$. Then, if
    $v_i(1) < -1$, then \cref{thr:value-adjustment-chores} allows us to rescale
    $i$'s valuations so that $v_i(1) = -1$, while maintaining that $v_i(M) \ge
    -|N|$ and $\mu_i \le -1$. Consequently, $I$ can be turned into an instance
    accepted by \cref{alg:2-MMS-chores} in polynomial time, and
    \cref{lem:correctness-chores} gurantees that a ($(2|N| -
    1)/|N|$)-approximate MMS allocation can be found in polynomial time.
    \qed
\end{proof}

\subsection{Proof for \cref{thr:3/2-MMS-chores-single-category}}

To prove \cref{thr:3/2-MMS-chores-single-category}, we need to develop a result
similar to \cref{lem:single-category-value-adjustment} and show that a similar
algorithm to \cref{alg:2/3-MMS} can in polynomial time find $3/2$-approximate
MMS allocations for a restricted class of instances. To simplify notation, we
assume that in any ordered instance, the chores are ordered so that the chore
numbered $1$ is the worst chore (provides most disutility) and the chore
numbered $|M|$ is the best chore (provides least disutility).

\begin{lemma}\label{lem:single-category-value-adjustment-chores}
    Let $I = \langle N, M, V, \langle (C_1, k_1) \rangle \rangle$ be an ordered
    instance of the the fair allocation of chores problem under cardinality
    constraints. Let $B_r = \{1, 2, \dots, r\}$ along with the $\max(0, |M| -
    (|N| - r)k_1 - r)$ best chores in $M$, for any $r \in \{1, 2, \dots, |N|\}$.
    Then, for any $i \in N$,
    \[
        v_i(B_r) \ge r\mu_i
    \]
\end{lemma}

\begin{proof}
    For any agent $i \in N$ and $r \in \{1, 2, \dots, |N|\}$, let $A$ be an MMS
    partition of $I$ for $i$. Let $A^*$ be the union of $r$ bundles in $A$ such
    that $\{1, 2, \dots, r\} \subseteq A^*$. At least one such $A^*$ must exist
    as the $r$ chores in $\{1, 2, \dots, r\}$ are contained in at most
    $r$ distinct bundles in $A$. Since $A$ is an MMS partition, we know that
    $v_i(A^*) \ge r\mu_i$. Since $A$ is feasible, $A^*$ must contain at least
    $\max(r, |M| - (|N| - r)k_1)$ chores.  Otherwise, there are more chores left
    in $M$ than can be contained in the $|N| - r$ bundles of $A$ not included in
    $A^*$. Since $B_r$ contains the chores $\{1, 2, \dots, r\}$ along with
    $\max(0, |M| - (|N| - r)k_1 - r)$ other chores, it follows that $|B_r| \le
    |A^*|$. Combined with the fact that the chores in $B_r \setminus \{1, 2,
    \dots, r\}$ are the chores in $M \setminus \{1, 2, \dots, r\}$ that provide
    the least disutility, we get that
    \begin{align*}
        v_i(B_r) &= v_i(\{1, 2, \dots, r\}) + v_i(B_r \setminus \{1, 2,
        \dots, r\}) \\
            &\ge v_i(\{1, 2, \dots, r\}) + v_i(A^* \setminus \{1, 2, \dots,
            r\}) \\
            &= v_i(A^*) \\
            &\ge r\mu_i
    \end{align*}
    \qed
\end{proof}

\noindent As was the case for goods,
\cref{lem:single-category-value-adjustment-chores} allows us to, in polynomial
time, scale valuations such that our estimates of $\mu_i$ achieves the required
accuracy for the bag-filling style algorithm (\cref{alg:3/2-MMS}). Additionally,
it provides a vital role in showing that for some agent $i \in N$, the bundle
created in \cref{alg:3/2-MMS} contains a sufficient number of chores when it is
worth more than $-3/2$.

With \cref{lem:single-category-value-adjustment-chores} we can now prove that
\cref{alg:3/2-MMS} finds a $(3/2)$-approximate MMS allocation for the instances
that fulfills the input requirements.

\begin{figure}[t]
    \begin{procedure}{Find $(3/2)$-MMS solution for single-category instance}{3/2-MMS}
        \Preamble{An ordered instance $I = \langle N, M, V, \langle C_1, k_1
        \rangle \rangle$ with $|M| > |N|$, $\mu_i \le -1$, $v_i(B_r) \ge
        -r$ (from \cref{lem:single-category-value-adjustment-chores}),
        $v_i(1) > -1$, and $v_i(|N| + 1) > -1/2$ for every $i \in N$, $r
        \in \{1, 2, \dots, |N|\}$}{Allocation $A$ consisting of each bundle $B'_j$
        allocated}
        \begin{pseudo}
            let $n = |N|$
            let $B'_1 = \{1\}, B'_2 = \{2\}, \dots, B'_{n} = \{n\}$ \\
            \kw{for} $j = 1$ up to $n$ \\+
                \kw{if} $|M| > |N|$ \\+
                    add the $\min(|M| - |N|, k_1 - 1)$ worst chores in
                    \label{line:3/2-MMS-initial-chores}\\+*
                  & $M \setminus (B'_j \cup B'_{j + 1} \cup \dots \cup
                  B'_n)$ to $B'_j$ \\--
                \kw{while} $v_i(B'_j) < -3/2$ for all $i \in N$ and $M
                \setminus (B'_{j} \cup B'_{j + 1} \cup \dots \cup B'_n)$
                contains \label{line:3/2-MMS-chore-upgrade}\\+++*
                & a better chore than the worst chore in $B'_j \setminus
                \{j\}$\\--
                    exchange the worst chore $g \in B'_j \setminus \{j\}$
                    for the worst chore \\+*
                    & $g' \in M \setminus (B'_j \cup B'_{j + 1} \cup \dots
                    \cup B'_n)$ with $g < g'$ \\--
                \kw{while} $v_i(B'_j) < -3/2$ for all $i \in N$
                \label{line:3/2-MMS-chore-dropping}\\+
                    remove the worst chore $g \in B'_j \setminus \{j\}$
                    from $B'_j$ \\-
                find $i \in N$ such that $v_i(B'_j) \ge -3/2$ \\
                allocate $B'_j$ to $i$ and set $N = N \setminus \{i\}$, $M
                = M \setminus B'_j$.
        \end{pseudo}
    \end{procedure}
\end{figure}

\begin{lemma}
    For an instance $I$ of the fair allocation of chores problem under
    cardinality constraints satisfying the requirements of \cref{alg:3/2-MMS},
    \cref{alg:3/2-MMS} finds a $(3/2)$-approximate MMS allocation in polynomial
    time.
\end{lemma}

\begin{proof}
    To show that the algorithm finds a $(3/2)$-approximate MMS allocation in
    polynomial time, we wish to show that
    \begin{inl}
        \item\label{item:3/2-MMS-condition-1}any $B'_j$ allocated is feasible,
        \item\label{item:3/2-MMS-condition-2}all chores are allocated,
        \item\label{item:3/2-MMS-condition-3}there is always an agent $i \in N$
            with $v_i(B'_j) \ge -3/2$, and
        \item\label{item:3/2-MMS-condition-4}the algorithm finishes in
            polynomial time.
    \end{inl}
    \\
    \vspace{0.5\baselineskip}

    \noindent \textbf{Feasible bundles}
    \\
    \vspace{0.2\baselineskip}

    \noindent To show \ref{item:3/2-MMS-condition-1} it suffice to show that
    $|B'_j| \le k_1$ at any point during the algorithm. Initially this holds, as
    the instance $I$ is assumed to have at least one feasible allocation and
    each chore must be allocated, hence $k_1 > 1$. Further, on
    line~\ref{line:3/2-MMS-initial-chores}, at most $k_1 - 1$ chores are added
    to $B'_j$. Since $|B'_j| = 1$ before this line, the size of $B'_j$ remains
    at most $k_1$. In the first loop (line~\ref{line:3/2-MMS-chore-upgrade}),
    the size of $B'_j$ does not change, as each iteration exchanges one good in
    $B'_j$ for one outside of $B'_j$. In the second loop
    (line~\ref{line:3/2-MMS-chore-dropping}), chores are removed from $B'_j$ and
    the size of $B'_j$ decreases. Hence, $|B'_j| \le k_1$, which guarantees that
    $B'_j$ is feasible when allocated and \ref{item:3/2-MMS-condition-1} holds.
    \\
    \vspace{0.5\baselineskip}

    \noindent \textbf{At the end of iteration $\mathbf{j}$, $\mathbf{v_i(B'_j)
    \ge -3/2}$ for an $\mathbf{i \in N}$}
    \\
    \vspace{0.2\baselineskip}

    \noindent As $v_i(g) > -1$ for all $i \in N$ and $g \in M$, we know that
    $v_i(\{j\}) \ge -1$. Since the second loop
    (line~\ref{line:3/2-MMS-chore-dropping}) removes one and one chore from
    $B'_j$, except for $j$, $B'_j$ must at some point be worth more than
    $-3/2$ to some agent $i \in N$. Otherwise, $B'_j$ would become $\{j\}$,
    which is worth at least $-1 > -3/2$ to all agents in $N$. Consequently,
    every bundle allocated is worth no less than $-3/2$ to the agent receiving
    it and a bundle is allocated in every iteration.
    \\
    \vspace{0.5\baselineskip}

    \noindent \textbf{All chores are allocated}
    \\
    \vspace{0.2\baselineskip}

    \noindent Showing that all of the chores are allocated boils down to showing
    that $B'_j$ contains a sufficiently large number of chores when allocated. To
    show that $B'_j$ has sufficient size, we will use a similar loop invariant
    argument as used to show that a sufficient amount of value remained for
    single-category instances of goods. Let $B_r^j$ denote
    the collection consisting of $\{j, j + 1, \dots, r\}$ and the $\max(0,
    |M| - (n - r)k_1 - (r - j + 1))$ best chores in $M$ at the start of
    iteration $j$. In other words, $B_r^j$ contains $\{j, j + 1, \dots,
    r\}$ and if $|M \setminus \{j, j + 1, \dots, r\}| > (n - r)k_1$,
    $B_r^j$ contains the exact number of chores needed so that
    $|M \setminus B_r^j| \le (n - r)k_1$. These additional chores are the
    best (least disutility) remaining chores. Note that if $|M| \le k_1|N|$
    after iteration $j - 1$, then by definition $|B_r^j| \le (r - j +
    1)k_1$.

    We wish to show that the two following properties hold for $B_r^j$ for
    all $r \in \{1, 2,
    \dots, n\}$ and $j \le n$:
    \begin{enumerate}
        \item\label{condition:3/2-MMS-1} $v_i(B_r^j) \ge -(r - j + 1)$ for
            all $i \in N$
        \item\label{condition:3/2-MMS-2} $|B_r^j| \le k_1(r - j + 1)$
    \end{enumerate}
    Specifically, this would mean that when $r = j$, then $v_i(B_r^j) \ge
    -1$ for all $i \in N$ and $|M \setminus B_r^j| \le |N \setminus
    \{i\}|k_1$. In other words, the bundle $B_r^j$ is such that if $B_r^j$ is
    allocated, then after allocation there remains at most as many chores as can
    be given to the remaining agents and the bundle may be given to any one of
    the agents without violating the MMS approximation guarantee. Especially,
    when $j = n$, this would mean that all remaining chores can and will be
    allocated to the remaining agent (line~\ref{line:3/2-MMS-initial-chores}
    will for $j = n$ add all chores in $M$ to $B'_j$ if $|M| \le k_1$).

    Notice how $B_r^j = B_r$ when $j = 1$. Consequently, by
    \cref{lem:single-category-value-adjustment-chores} we have $v_i(B_r^1)
    \ge -r$. Additionally, since $|M| \le nk_1$ at the start,
    \[
        |B_r^1| = r + \max(|M| - (n - r)k_1 - r, 0) \le r + nk_1
        - (n - r)k_1 - r = r k_1
    \]

    \noindent Thus, the conditions hold for $j = 1$. Assume for some $j < n$
    that they hold for all $r \ge j$. We wish to show that they hold for $j'
    = j + 1$ and all $r \ge j'$. For \ref{condition:3/2-MMS-2}., notice that
    if $B'_j$ is not modified in the second loop
    (line \ref{line:3/2-MMS-chore-dropping}), then either $|B'_j| = |M| - |N| +
    1$ and $B'_j$ contains all chores in $M \setminus \{j + 1, j + 2, \dots,
    n\}$, or $|B'_j| = k_1$. In the first case, $|M \setminus B'_j| = |\{j', j' + 1,
    \dots, r\}| = r - j' + 1 \le k_1|N \setminus \{i\}|$. In the second
    case, since $|B_j^j| \le k_1$, $|M| \le |N|k_1$ and $|M \setminus B'_j| \le
    |N \setminus \{i\}|k_1$. Consequently, we know $|B_r^{j'}| \le k_1(r -
    j' + 1)$ in either case.

    If $B'_j$ is modified in the second loop, then note that before the first
    iteration of the loop, $B'_j$ consists of $j$ and the $\min(|M| - |N|, k_1 -
    1)$ best chores in $M$. This is the exact same construction as
    $\smash{B_j^j}$, except that $\smash{B_j^j}$ contains $|\smash{B_j^j}| - 1
    \le \min(|M| - |N|, k_1 - 1)$ of the best chores in $M$. Since the loop
    removes the worst chore in $B'_j \setminus \{j\}$ in each iteration, $B'_j$
    will turn into $\smash{B_j^j}$ at some point. Since $v_i(\smash{B_j^j}) \ge
    -1$, $|B'_j| \ge |\smash{B_j^j}|$ when the second loop finishes. By
    definition, $|M \setminus B'_j| \le |M \setminus \smash{B_j^j}| \le k_1(n -
    j) = k_1(|N \setminus \{i\})$ and \ref{condition:3/2-MMS-2}.\ holds in all
    cases.

    For \ref{condition:3/2-MMS-1}.\ first note that any change performed in the
    first or the second loop (lines~\ref{line:3/2-MMS-chore-upgrade} and
    \ref{line:3/2-MMS-chore-dropping}) either removes a chore in $M
    \setminus \{1, 2, \dots, n\}$ from $B'_j$ or exchanges a chore from that
    subset of $M$ for another in the same subset. That is, the value of $B'_j$
    changes by at most $1/2$ in each operation. Thus, either $v_{i'}(B'_j) < -1$
    for all $i' \in N$ or $B'_j$ is not modified in either loop.

    If $B'_j$ is modified in the last loop
    (line~\ref{line:3/2-MMS-chore-dropping}), then since $\smash{B_r^j}
    \setminus \{j, j + 1, \dots, r\}$ and $B'_j \setminus \{j\}$ both consist of
    some number of the best chores in $M$, either $B'_j \subseteq \smash{B_r^j}$
    or $(\smash{B_r^j} \setminus \{j + 1, j + 2, \dots, r\}) \subseteq B'_j$. In
    the first case,
    \[
        v_{i'}(B_r^{j'}) = v_{i'}(B_r^j) - v_{i'}(B'_j) \ge -(r - j +
        1) - (-1) = -(r - j' + 1)
    \]
    \noindent for all $i' \in N \setminus \{i\}$. In the second case,
    $\smash{B_r^{j'}} = \{j', j' + 1, \dots, r\}$ and $v_{i'}(\smash{B_r^{j'}})
    \ge -(r - j' + 1)$ since $v_{i'}(g) \ge - 1$ for $g \in M$.

    If $B'_j$ is not modified in the second loop, then we
    know that either $|B'_j| < k_1$ and $M \setminus B'_j = \{j', j' + 1, \dots
    n\}$ or $|B'_j| = k_1$. In the first case, we have as earlier
    $\smash{B_r^{j'}} = \{j', j' + 1, \dots r\}$ and $v_i(\smash{B_r^{j'}}) \ge
    -(r - j' + 1)$. In the second case, if $|B_r^{j}| < k_1 - 1 + (r - j + 1)$,
    then $\smash{B_r^{j'}} = \{j', j' + 1, \dots, n\}$ and
    $v_{i'}(\smash{B_r^{j'}}) \ge -(r - j' + 1)$.  Otherwise, we know that by
    the way the chores are exchanged in the loop, we always select the worst $g'
    \in M \setminus (B'_j \cup B'_{j + 1} \cup \dots \cup B'_{n})$ that is
    better than the chore replaced. In other words, there is no chore in $M
    \setminus B'_j$ that is both better than and worse than two distinct chores
    in $B'_j \setminus \{j\}$. Thus, either $B'_j \subseteq \smash{B_r^{j}}$ or
    there is no chore $g \in B'_j$ such that there is $g' \in \smash{B_r^{j}}
    \setminus \{j, j + 1, \dots, r\}$ with $g' < g$. In other words, $B'_j
    \setminus \{j\}$ contains $k_1 - 1$ chores such that for $g \in B'_j
    \setminus \{j\}$, $g$ is either worse than the chores in $\smash{B_r^j}
    \setminus \{j, j + 1, \dots, r\}$ or $B'_j$ contains all worse chores in
    $\smash{B_r^j} \setminus \{j, j + 1, \dots, r\}$. Consequently, since
    $|\smash{B_r^j}| \le k_1(r - j + 1)$, the $k_1$ chores removed from
    $\smash{B_r^j}$ to create $\smash{B_r^{j'}}$ are $j$ and the $k_1 - 1$
    worst chores in $\smash{B_r^j} \setminus \{j, j + 1, \dots, r\}$.  Due to
    the ordered instance, these $k_1$ chores must be at least $1/(r - j + 1)$ of
    the disutility in $\smash{B_r^j}$. Consequently,
    \[
        v_{i'}(B_r^{j'}) \ge \left(1 - \frac{1}{r - j +
        1}\right)v_{i'}(B_r^{j}) \ge \frac{r - j}{r - j + 1} \cdot
        -(r - j + 1) = -(r - j' + 1)
    \]
    \noindent for all $i' \in N \setminus \{i\}$. Consequently,
    \ref{condition:3/2-MMS-1}.\ holds and it follows that
    \ref{item:3/2-MMS-condition-2} holds.
    \\
    \vspace{0.5\baselineskip}

    \noindent \textbf{Polynomial run time}
    \\
    \vspace{0.2\baselineskip}

    \noindent It remains to show that \ref{item:3/2-MMS-condition-4} holds,
    namely that the algorithm run in polynomial time. Since
    \ref{item:3/2-MMS-condition-1}, \ref{item:3/2-MMS-condition-2} and
    \ref{item:3/2-MMS-condition-3} hold, it follows that the algorithm will not
    be stuck in any loop without any operations to perform. Further, since $j$
    is bounded in the number of agents, it suffice to show that each iteration
    of the outer loop can be performed in polynomial time. Since each iteration
    of the first loop improves the worst chore in $B'_j$, the number of
    iterations of this loop is at most $|M|$. In the second loop, a chore is
    removed from $B'_j$ in each iteration. Consequently, the loop can at most
    have $|B'_j \setminus \{j\}| \le |M|$ iterations. Since each of the
    individual operations can be performed in polynomial time, it therefore
    follows that each iteration of the outer loop can be performed in polynomial
    time and \ref{item:3/2-MMS-condition-4} holds.
    \qed
\end{proof}

\begin{proof}[for \cref{thr:3/2-MMS-chores-single-category}]
    To show that $3/2$-approx\-imate MMS allocations can be found in polynomial
    time, we will show that in polynomial time, $I$ can either be
    converted into an instance that \cref{alg:3/2-MMS} accepts or a
    $3/2$-approximate MMS allocation can trivially be found. First, note that if
    $v_i(M) = 0$, for some agent $i \in N$, then agent $i$ can be allocated the
    worst $k_1$ chores, as agent $i$ assigns each of these a value of $0$. The
    instance without $i$ and these chores is obviously feasible and one in which
    the MMS of each agent is no worse than in $I$. Thus, any $i$ with $v_i(M) =
    0$ can be reduce away, and we can assume that $v_i(M) < 0$ for each
    remaining agent $i \in N$. If $|M| \le |N|$, then any allocation that gives
    each agent at most one chore is an MMS allocation. Such an allocation can
    trivially be found in linear time.

    If $|M| > |N|$, the only change needed for $I$ is to rescale the valuations
    of the agents. First, by \cref{thr:normalization-chores}, we can for each
    agent $i \in N$ rescale $i$'s valuations so that $v_i(M) = |N|$ which
    guarantees $\mu_i \le -1$. Further, by \cref{thr:value-adjustment-chores}
    if $v_i(1) < -1$, then adjusting $i$'s valuations so that $v_i(1) = -1$
    maintains $\mu_i \le -1$. If $v_i(|N| + 1) < -1/2$,
    \cref{thr:value-adjustment-chores} also guarantees that $\mu_i \le -1$ if we
    rescale $i$'s valuations so that $v_i(|N| + 1) = -1/2$. Similarly, by
    \cref{lem:single-category-value-adjustment-chores} if $v_i(B_r) < -r$
    for some $r \in \{1, 2, \dots, |N|\}$, then $i$'s valuations can be
    rescaled so that $v_i(B_r) = -r$ while still guaranteeing $\mu_i \le
    -1$. Since each rescale increases $v_i(M)$ and all the properties required
    for \cref{alg:3/2-MMS} require the value of a set of chores to be above a
    certain threshold, each further rescale does not break any
    one of the properties that already hold. Thus, after checking all the cases
    above, all conditions of \cref{alg:3/2-MMS} hold. Since both $r$ and $i$
    are bound in the number of agents, the conversion can be performed in
    polynomial time.
    \qed
\end{proof}

\section{Omitted Examples}

\begin{example}[Failing valid reduction with two goods]
    \label{exm:failing-reduction}
    As mentioned in \cref{sec:reduced-instances}, even for instances with a
    single category, certain types of valid reductions from unconstrained fair
    allocation fail to be applicable. One of these is the valid reduction
    created by constructing a bundle $B$ consisting of two goods such that
    $v_i(B) \ge \alpha\mu_i$ for an agent $i$ and $v_{i'} \le \mu_{i'}$ for all
    other agents $i'$. In unconstrained fair allocation, this is a valid
    reduction since for any agent $i'$, one can easily show that their MMS
    remains at least as high. This follows from the fact that the goods in
    $B$ are contained in either one or two bundle in the agent's MMS partition.
    Therefore, there is either already $n - 1$ bundles valued at $\mu_{i'}$ or
    higher, or the two bundles containing goods from $B$ can be combined to have
    a value of at least $\mu_{i'}$. Under cardinality constraints, this last
    step could lead to infeasibility, and the reduction therefore does not work,
    as seen below.

    Let $I$ be an instance of the fair allocation problem under cardinality
    constraints, with a single category $C_1$ with threshold $k_1 =
    5$, $11$ goods and $3$ agents with identical valuation functions given in
    \cref{tab:failing-reduction}. The MMS of each agent is $1$, as $v_i(M) = 3$
    and the partition $(\langle 1, 8, 9 \rangle, \langle 2, 10, 11 \rangle,
    \langle 3, 4, 5, 6, 7 \rangle)$ contains three bundles with a value of
    exactly $1$.  If $\alpha \le 19/20$, then the bundle $B = \langle 2, 7
    \rangle$ would constitute a valid reduction for any agent in the
    unconstrained setting.  However, with the given threshold for $C_1$,
    removing $B$ and an agent $i$ would leave us with an instance where the MMS
    of the agents would be $37/40$ with the MMS partition $(\langle 1, 8, 9, 10
    \rangle, \langle 3, 4, 5, 6, 11 \rangle)$. If $37/40 < \alpha \le 19/20$,
    this would even make it impossible to achieve an $\alpha$-approximate MMS
    allocation.

    \begin{table}[H]
        \centering
        \begin{tabular}{cccccccccccc}
            \toprule
            $j$ & 1 & 2 & 3 & 4 & 5 & 6 & 7 & 8 & 9 & 10 & 11 \\
            \midrule
            $v_i$ & $\frac{3}{4}$ & $\frac{3}{4}$ & $\frac{1}{5}$ &
            $\frac{1}{5}$ & $\frac{1}{5}$ & $\frac{1}{5}$ & $\frac{1}{5}$ &
            $\frac{1}{8}$ & $\frac{1}{8}$ & $\frac{1}{8}$ & $\frac{1}{8}$ \\
            \bottomrule
            \vspace{0.2em}
        \end{tabular}
        \caption{Valuations in \cref{exm:failing-reduction}}
        \label{tab:failing-reduction}
    \end{table}
\end{example}

\end{document}